\documentclass[toc]{PoS}
\usepackage{amsmath,amsfonts,amssymb,amsthm}
\usepackage{url}
\usepackage{array}

\title{Introduction to Black Hole Evaporation}

\ShortTitle{Introduction to Black Hole Evaporation}

\author{\speaker{Pierre-Henry Lambert}\\
          Physique Th\'eorique et Math\'ematique \\ Universit\'e Libre de
   Bruxelles and International Solvay Institutes \\ Campus
   Plaine C.P. 231, B-1050 Bruxelles, Belgium\\
        E-mail: \email{pilamber@ulb.ac.be}}

\abstract{These lecture notes are an elementary and pedagogical introduction to the black hole evaporation, based on a lecture
given by the author at the Ninth Modave Summer School in Mathematical Physics and are intended for PhD students.\\ 
First, quantum field theory in curved spacetime is studied and tools needed for the remaining of the course are introduced.
Then, quantum field theory in Rindler spacetime in 1+1 dimensions and in the spacetime of a spherically collapsing star are considered, leading to Unruh and Hawking effects, respectively.
Finally, some consequences such as thermodynamics of black holes and information loss paradox are discussed.}

\FullConference{Ninth Modave Summer School in Mathematical Physics,\\
		1-7 September, 2013\\
		Modave, Belgium}

\graphicspath{{endimages/}}
\begin{document}

\section*{Introduction}
\addcontentsline{toc}{section}{Introduction}
%------------------------
%------------------------

There are several reasons for which the general relativity theory cannot be the final theory in order to describe the gravitational interaction. From one side the theory predicts the existence of singularities but does not resolve them, this is clearly an internal evidence that general relativity theory is incomplete. On the other side, general relativity is a classical theory but Nature is fundamentally quantum (in a certain sense) and at present a quantum theory of gravity is still missing. This is an external evidence that general relativity cannot be the final theory concerning gravity.

Even though a theory of quantum gravity is not available, one can nevertheless try to gain some information about the quantum properties of gravity by using an approximation scheme. A scheme one can consider is the semi-classical theory in which the gravity is treated classically but on which quantum fields can propagate. In this approximation scheme, the quantum fields satisfy their usual equations of motion but with the usual Minkowski constant metric replaced by the classical metric of the curved spacetime under consideration.

The main aim of this lecture notes is to present, in a very basic and (hopefully) pedagogical way, the surprising result discovered by Hawking \cite{HawkingP} in which black holes are shown to create and emit particles in the semi-classical approach when quantum effects are taken into account, contrary to the prediction of the classical theory. These notes are based on a lecture that was given by the author at the Ninth Modave Summer School in Mathematical Physics and was intended for PhD students that are not necessarily familiar with quantum field theory in curved spacetimes. 
Emphasis is put in such a way of providing a very pedestrian and elementary introduction about this topic with no other ambition than being as self-contained as possible. Some standard notions of general relativity are supposed to be known, in particular the notions of black holes and Carter-Penrose diagrams.
These lecture notes are so pitched that graduate students familiar with these notions should have essentially no difficulty in following them.

In practice, these notes are mainly based on the original paper of Hawking \cite{HawkingP}, on the general relativity book of Carroll \cite{Carroll} and on the notes on black holes of Townsend \cite{Townsend} and Dowker \cite{Dowker}. Interested readers are warmly invited to consult some other references such as the book of Birrell and Davies \cite{Birrell} from the eighties, or the one of Wald \cite{Wald} from the nineties or the more recent one of Mukhanov and Winitski \cite{Mukhanov}.

These lecture notes are organized as follow. In the first section, general properties of quantum field theory in curved spacetime are studied and differences with respect to the flat space are pointed out. Applications of the general formalism are then considered. First in the case of a sandwich spacetime before considering the more physically interesting case of Rindler space in 1+1 dimensions and also the spacetime of a spherically collapsing star. This will lead to Unruh and Hawking effects, respectively. Finally some consequences of the result of Hawking are discussed, namely the thermodynamics of black holes and the information loss paradox. 
Several figures are included, in order to make the presentation of the topic more pedagogical.

Any comments about these lecture notes are welcome.

\section{Quantum Field Theory in curved spacetime}
%---------------------------------------------------------------
%---------------------------------------------------------------

\subsection{Quick review of QFT in flat spacetime}
%--------------------------------------------------------------
%This part is mainly based on \cite{Carroll} and \cite{Dowker}.
The signature is taken to be $(-,+,+,+)$ throughout the lecture. Let us consider the following action \cite{Carroll}\cite{Dowker}
\begin{eqnarray}
S=\int d^4x ~\mathcal L,\hspace{1cm}\text{with}\hspace{1cm}\mathcal L=-\frac{1}{2}\partial_\mu\phi\partial^\mu\phi-\frac{1}{2}m^2\phi^2.
\end{eqnarray}
The equation of motion, obtained by requiring the action to be stationary, reads
\begin{eqnarray}\label{1.2}
\delta S=0\hspace{1cm}\Rightarrow\hspace{1cm} \Box\phi-m^2\phi=0,
\end{eqnarray}
where $\Box=\partial_\mu\partial^\mu$. Equation \eqref{1.2} is the familiar Klein-Gordon equation.
The conjugate momentum is defined by $\pi=\frac{\partial \mathcal L}{\partial\partial_0\phi}=\dot\phi$, where $\dot\phi=\partial_t\phi$.

\noindent A set of solutions to the Klein-Gordon equation of motion is given by plane waves,
\begin{eqnarray}\label{1.3}
f=f_0e^{ik_\mu x^\mu},
\end{eqnarray}
with $k^\mu=(\omega,k^i), k_\mu=(-\omega,k_i)$ where $\omega$ is the frequency and $k^i$ the wave vector. In the rest of the lecture, the wave vector $k^i$ will be denoted by $k$, except when explicitely stated. The dispersion relation is obtained by replacing the plane wave solution \eqref{1.3} into the Klein-Gordon equation \eqref{1.2},
\begin{eqnarray}
\omega^2=k^2+m^2,\label{1.4}
\end{eqnarray}
with $k^2=k_ik^i$. This relation means that the wave vector $k$ completely determines the frequency $\omega$, up to a sign. The frequency $\omega$ is chosen to be a positive number and so the set of solutions \eqref{1.3} becomes parameterized by the wave vector $k$:
\begin{eqnarray}\label{1.5}
f_k=f_0e^{ik_\mu x^\mu}.
\end{eqnarray}
By definition, modes $\{f_k\}$ such that
\begin{eqnarray}
\partial_t f_k=-i\omega f_k, \hspace{1cm}\text{with}\hspace{1cm}\omega>0,
\end{eqnarray}
are called positive frequency modes. Similarly, modes $\{f_k ^*\}$ such that
\begin{eqnarray}
\partial_t f_k^*=i\omega f_k^*,\hspace{1cm}\text{with}\hspace{1cm}\omega>0,
\end{eqnarray}
are called negative frequency modes (even though $\omega$ is positive). One can ask why the positive modes are written like this and the answer is that  it will be easier to generalize this notion later on when quantum field theory will be considered in curved spacetime.

In order to have a complete and orthonormal set of modes, an inner product must be defined on the space of solutions of the Klein-Gordon equation of motion. The inner product between two solutions $f$ an $g$ is defined by
\begin{eqnarray}
(f,g)=-i\int d^3x~~(f\partial_t g^*-g^*\partial_t f).
\end{eqnarray}
The normalization of the set of modes is obtained by considering the inner product between two plane waves $f_{k_1}=e^{-i\omega_1 t+ik_1x}$ and $f_{k_2}=e^{-i\omega_2 t+ik_2x}$,
\begin{align}
(f_{k_1},f_{k_2})&=-i\int d^3x~~i(\omega_1+\omega_2)e^{-i(\omega_1-\omega_2)t}e^{i(k_1-k_2)x},\notag\\
&=(\omega_1+\omega_2)~\delta^3(k_1-k_2)(2\pi)^3e^{-i(\omega_1-\omega_2)t},\label{2.10}
\end{align}
where the definition of the delta distribution was used, i.e. $\delta^3(k_1-k_2)=\int\frac{d^3x}{(2\pi)^3}e^{i(k_1-k_2)x}$.
From \eqref{2.10} one can see that the inner product between two different sets of modes vanishes unless the two wave vectors $k_1,k_2$ (and thus the corresponding frequency, by the dispersion relation) are equal. The set of modes \eqref{1.5} can then be normalized as
\begin{eqnarray}
f_k=\frac{1}{(2\pi)^{3/2}}\frac{1}{\sqrt{2\omega}}e^{ik_\mu x^\mu}.
\end{eqnarray}
This normalization implies the relations
\begin{eqnarray}
(f_k,f_{k^\prime})=\delta(k-k^\prime),\hspace{1cm}(f_k,f_{k^\prime}^*)=0,\hspace{1cm}(f_k^*,f_{k^\prime}^*)=-\delta(k-k^\prime).
\end{eqnarray}
Any classical field configuration $\phi(x)$ that is a solution to the Klein-Gordon equation can be expanded in terms of the basis modes $\{f,f^*\}$,
\begin{eqnarray}
\phi(x)=\int d^3k ~~(a_kf_k+a_k^*f_k^*),
\end{eqnarray}
where $a_k$ and $a_k^*$ are some coefficients with respect to the basis modes in the expansion of the field configuration.

The scalar field can be canonically quantized, by replacing the classical fields by operators acting on Hilbert space and by imposing the canonical commutation relation
\begin{eqnarray}
[\phi(t,x),\pi(t,x)]=\delta(x-x^\prime).\label{2.14}
\end{eqnarray}
After quantization, the classical field $\phi$ expanded in terms of modes becomes the following operator
\begin{eqnarray}
\phi=\int d^3k~~( a_k f_k+a_k^\dagger f_k^*).\label{2.15}
\end{eqnarray}
Note that the operator $a_k$ in the expansion of the quantum field $\phi$ was defined at the classical level to be the coefficient of the positive frequency mode. This fact seems anecdotic right now but will however play an important role later on when quantum field theory will be considered in curved spacetime. From the commutation relation \eqref{2.14} and the field expansion \eqref{2.15} one can easily check that the operators $a_k, a_k^\dagger$ satisfy
\begin{eqnarray}
[a_k,a_{k^\prime}^\dagger]=\delta(k-k^\prime)\label{2.16}.
\end{eqnarray}
Relation \eqref{2.16} is exactly the same relation as the one for the creation/annihilation operators of the familiar quantum harmonic oscillator, except that there is one oscillator for each mode $k$. As for the harmonic oscillator, the vacuum is defined to be the state $|0\rangle$ such that $a_k|0\rangle=0,~~ \forall k$, and the number of particles of momentum $k$ is defined by operator $N_k=a_k^\dagger a_k$, with no $k$ summation in the right hand side.

\subsection{QFT in curved spacetime}
%--------------------------------------------
Let us move to curved spacetime by choosing some background $(M,g)$. Suppose that $(M,g)$ is globally hyperbolic. What does that mean?

\subsubsection{Global hyperbolicity}
In this section, based on \cite{Townsend}, \cite{Dowker}, \cite{Henneaux}, \cite{Ellis}, \cite{Isham} and \cite{Geroch} the notion of global hyperbolicity is considered. First some definitions are needed.\\

\noindent \textsc{Definition 1}\\
The future domain of dependence of a hypersurface $\Sigma$, denoted by $D^+(\Sigma)$, is the set of points $p\in M$ for which every past (inextendable) causal curves through $p$ intersect $\Sigma$.\\

\begin{figure}[!th]
       \centering
\begin{minipage}[t]{7.5cm}
\centering
      \includegraphics[width=7.5cm]{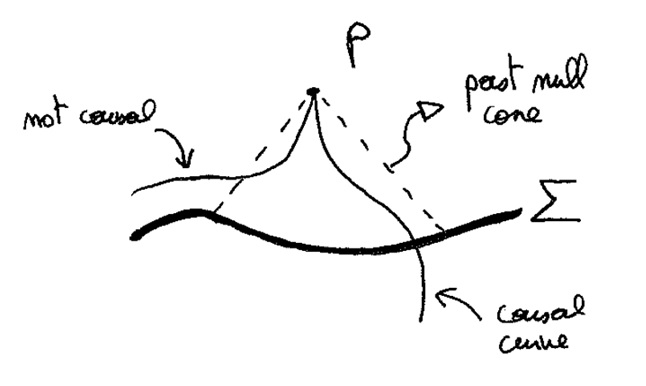}
     \caption{Past causal curve throught point $p$.}
    \label{fig1}
\end{minipage}
\begin{minipage}[t]{7.5cm}
\centering
      \includegraphics[width=5cm]{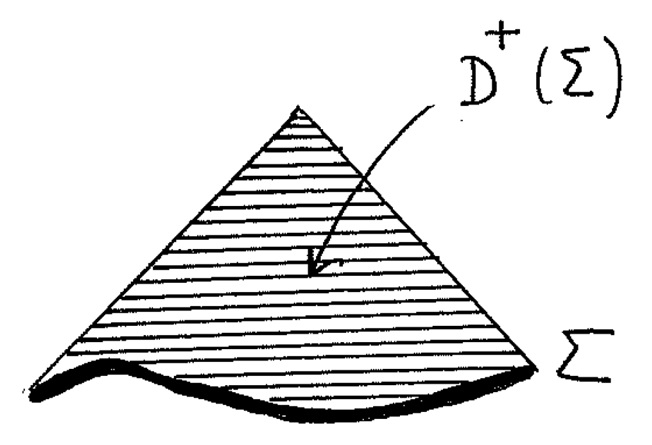}
     \caption{The future domain of dependence of $\Sigma$.}
    \label{fig1bis}
\end{minipage}
\end{figure}

This definition is illustrated by figures \ref{fig1} and \ref{fig1bis}.
\\
In the case of partial differential equation, the meaning of this notion is that the behavior of the solution of the partial differential equation outside $D^+(\Sigma)$ is not determined by initial data on $\Sigma$.\\

\noindent \textsc{Definition 2}\\
The past domain of dependence of a hypersurface $\Sigma$, denoted by $D^-(\Sigma)$, is the set of points $p\in M$ for which every future (inextendable) causal curves through $p$ intersect $\Sigma$.\\

\noindent \textsc{Definition 3}\\
$\Sigma$ is said to be a Cauchy surface for $(M,g)$ if $D^+(\Sigma)\bigcup D^-(\Sigma)=M$.\\

\noindent \textsc{Definition 4}\\
$(M,g)$ is globally hyperbolic if there exists (at least) one Cauchy surface $\Sigma$ for $(M,g)$.\\

\noindent Let us consider two examples of globally hyperbolic spacetimes.
\begin{enumerate}
\item Minkowski spacetime:
the Carter-Penrose diagram of Minkowski spacetime is given by figure \ref{fig2}.
Surfaces $\Sigma_1$ and $\Sigma_2$ are both Cauchy surfaces, because for any point inside the spacetime, every causal curve through it intersects both $\Sigma_1$ and $\Sigma_2$.
\begin{figure}[!th]
       \centering
\begin{minipage}[t]{7cm}
\centering
      \includegraphics[width=5cm]{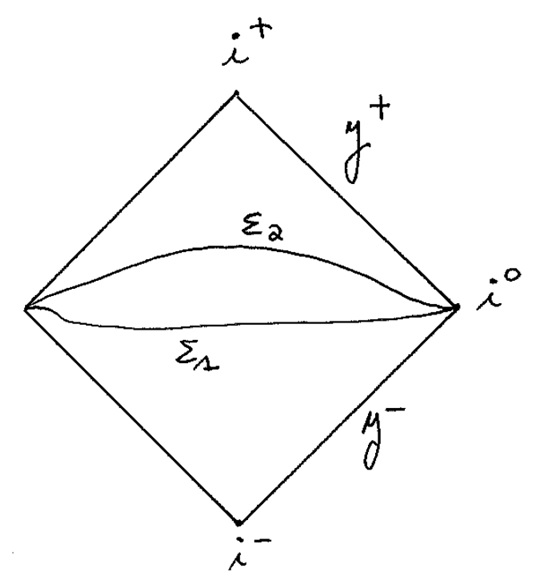}
     \caption{Carter-Penrose diagram for Minkowski.}
    \label{fig2}
\end{minipage}
\begin{minipage}[t]{2.2cm}
\end{minipage}
\begin{minipage}[t]{7cm}
\centering
      \includegraphics[width=6cm]{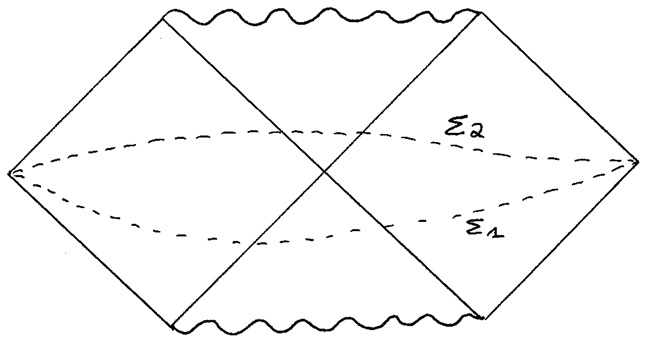}
     \caption{Carter-Penrose diagram for Kruskal.}
    \label{fig3}
\end{minipage}
\end{figure}
\item Kruskal spacetime (the maximal analytic extension of Schwarzschild):
The Carter-Penrose diagram of Kruskal spacetime is given by figure \ref{fig3}. Again $\Sigma_1$ and $\Sigma_2$ are Cauchy surfaces.
\end{enumerate}

\noindent In the case where the spacetime $(M,g)$ is not globally hyperbolic, then either $D^+(\Sigma)$ or $D^-(\Sigma)$ has a boundary on $M$.\\

\noindent \textsc{Definition 5}\\
The future/past Cauchy horizon is the boundary of $D^+(\Sigma)~/~D^-(\Sigma)$ in $M$.\\

\noindent Let us consider two examples of non-globally hyperbolic spacetimes and of Cauchy horizons.
\begin{figure}[!th]
       \centering
\begin{minipage}[t]{7.5cm}
\centering
      \includegraphics[width=7.5cm]{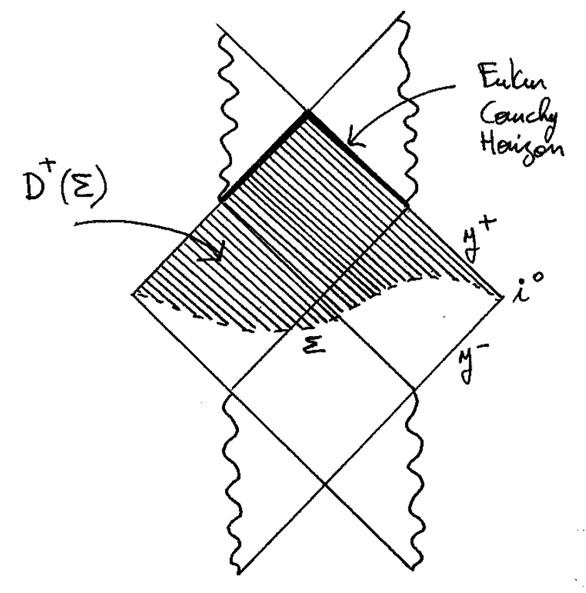}
     \caption{Future Cauchy horizon in Reissner-Nordstrom spacetime.}
    \label{fig5}
\end{minipage}
\begin{minipage}[t]{7.5cm}
\centering
      \includegraphics[width=6cm]{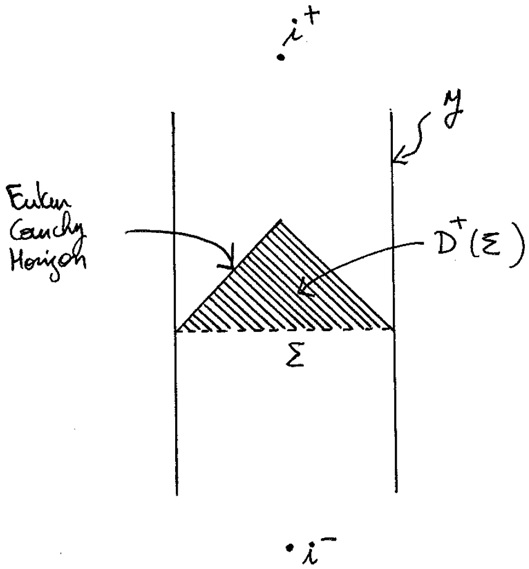}
     \caption{Carter-Penrose diagram for AdS. Presence of future Cauchy horizon.}
    \label{fig4}
\end{minipage}
\end{figure}
\begin{enumerate}
\item Maximal analytic extension of Reissner-Nordstrom:
the Carter-Penrose diagram for this spacetime is given by figure \ref{fig5}.
The aim of this example is not to enter into the various (interesting) details of this spacetime, but instead to illustrate the notion of Cauchy horizon for non-globally hyperbolic spacetime.

\item Anti de Sitter spacetime.
In AdS spacetime, null and spacelike infinity can be thought as timelike surfaces \cite{Ellis}, as can be seen from the Carter-Penrose diagram given by figure \ref{fig4}. This is an indication that the spacetime is not globally hyperbolic. 
Moreover, any spacelike surface (like $\Sigma$ in figure \ref{fig4}) has a Cauchy horizon, this fact proves that AdS is not globally hyperbolic.
To get more feeling about the non-globally hyperbolicity of AdS let us consider null geodesics in AdS$_2$, for simplicity
\begin{align}
ds^2&=-\cosh^2 r ~dt^2+dr^2\notag\\
ds^2&=0\hspace{1cm}\Rightarrow t=\pm\int\frac{dr}{\cosh r}=\pm\int2\frac{e^rdr}{e^{2r}+1}=\pm 2\arctan e^r.\notag
\end{align}
\begin{figure}[!th]
       \centering
      \includegraphics[width=5cm]{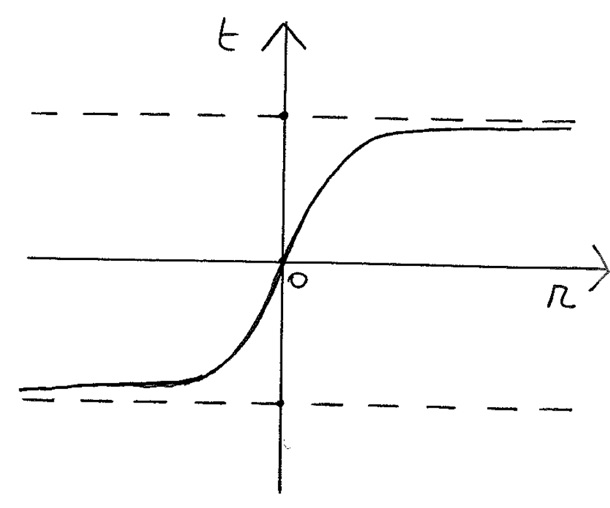}
     \caption{Lightlike geodesics in AdS$_2$.}
    \label{fig6}
\end{figure}
A photon can thus reach $r=\infty$ in a finite time, see figure \ref{fig6}. Conversely, information can come from $r=\infty$ to some $r=r_0$ in a finite time. The fact that AdS is not globally hyperbolic means that if one wants to predict physics in AdS, initial conditions given on a spacelike hypersurface $\Sigma$ is not enough, initial conditions must also be given at infinity.
\end{enumerate}

\noindent However, doing quantum field theory in a non-globally hyperbolic spacetime is still possible but is more difficult, see \cite{Isham} for instance. Therefore in the rest of this lecture attention will be restricted only to globally hyperbolic spacetimes.

\subsubsection{QFT in globally hyperbolic spacetime}
%----------------------------------------------------------
For a background $(M,g)$ that is globally hyperbolic one can study a scalar field $\phi$ propagating on it, by considering the action
\begin{eqnarray}
S=\int d^4x \sqrt{-g}\left(-\frac{1}{2}D_\mu\phi D^\mu\phi-\frac{1}{2}m^2\phi^2\right),
\end{eqnarray}
where $D_\mu$ denotes the covariant derivative, and $D_\mu\phi=\partial_\mu\phi$ in the case of a scalar field $\phi$. The equation of motion, obtained by varying the action, reads
\begin{eqnarray}
\delta S=0\hspace{1cm}\Rightarrow\hspace{1cm}\frac{1}{\sqrt{-g}}\partial_\mu(\sqrt{-g}g^{\mu\nu}\partial_\nu)\phi-m^2\phi\equiv\Box\phi-m^2\phi=0,\label{2.19}
\end{eqnarray}
where $\Box\phi$ is defined to  be the first term in the left hand side of \eqref{2.19}. This is the usual Klein-Gordon equation.\\
The inner product on solutions space of the Klein-Gordon equation is defined by
\begin{eqnarray}\label{2.20}
(\phi_1,\phi_2)=-i\int_{\Sigma}d^3x~~\sqrt{\gamma}~n^\mu~~(\phi_1D_\mu\phi_2^*-\phi_2^*D_\mu\phi_1),
\end{eqnarray}
where $\Sigma$ is a Cauchy surface\footnote{This is the place where global hyperbolicity hypothesis is important.}, with normal vector $n^\mu$ and induced metric $\gamma$. This inner product is natural, i.e. is independent of the choice of the Cauchy surface $\Sigma$. Indeed, by considering two different Cauchy surfaces $\Sigma_1$ and $\Sigma_2$, we have
\begin{align*}
(\phi_1,\phi_2)|_{\Sigma_1}-(\phi_1,\phi_2)|_{\Sigma_2}&=-i\int_{\Omega=\Sigma_1-\Sigma_2}d^3x~~\sqrt{\gamma}~n^\mu~~(\phi_1D_\mu\phi_2^*-\phi_2^*D_\mu\phi_1),\\
&=-i\int_{\partial\Omega}d^4x~~\sqrt{-g}~D^\mu(\phi_1D_\mu\phi_2^*-\phi_2^*D_\mu\phi_1),\\
&=-i\int_{\partial\Omega}d^4x~~\sqrt{-g}~(\phi_1m^2\phi_2^*-\phi_2^*m^2\phi_1)=0,
\end{align*}
where Stokes theorem and the equation of motion were used to get the second and third lines, respectively. This inner product allows to have an (a priori non-unique) orthonormal basis satisfying
\begin{eqnarray}
(f_i,f_j)=\delta_{ij},\hspace{1cm}(f_i^*,f_j^*)=-\delta_{ij},
\end{eqnarray}
where indices $i,j,\dots$ can be discrete or continuous. In order to make things as easy as possible, the notation for the discrete case is adopted. This inner product allows to define an orthonormal basis, but this basis is non-unique and after canonical quantization of the theory there will be different notions of vacuum, according to each different orthonormal basis. This is because there is no preferred time coordinate in curved spacetime, except if another assumption is made.

\subsubsection{Stationary spacetime}
%---------------------------------------------
The assumption that is made on the spacetime $(M,g)$ is stationary symmetry.\\

\noindent \textsc{Definition 6}\\
A spacetime $(M,g)$ is stationary if there exists a timelike killing vector field $K=K^\mu\partial_\mu$ for the metric $g$, i.e. $\mathcal L_Kg=0$.\\

This definition implies that there exists a coordinate system $\{x^\mu\}$ such that the metric is time independent, i.e. $\partial_t g_{\mu\nu}=0$.
Indeed, in an arbitrary coordinate system the Killing equation is
\begin{eqnarray}
0=\mathcal L_Kg_{\mu\nu}\equiv K^\sigma\partial_\sigma g_{\mu\nu}+g_{\sigma\nu}\partial_\mu K^\sigma+g_{\mu\sigma}\partial_\nu K^\sigma,\label{2.24}
\end{eqnarray}
and choosing a coordinate system $\{x^\mu\}$ in which $K^\mu=(1,0,0,0)$ reduces \eqref{2.24} to $\partial_tg_{\mu\nu}=0$, hence the result. Roughly speaking, or at least intuitively, stationary means time independent.

So far so good. The globally hyperbolic background $(M,g)$ is stationary, with a Killing vector field $K$. This Killing vector field enjoys two properties.
\begin{enumerate}
\item First of all, it commutes with the Klein-Gordon operator $\Box-m^2$. Indeed, from one side we have
\begin{eqnarray*}
\partial_t\Box\phi=\partial_t (D_\mu g^{\mu\nu} \partial_\nu \phi)=g^{\mu\nu}\partial_t(\partial_\mu\partial_\nu\phi-\Gamma^\sigma_{\mu\nu}\partial_\sigma \phi)=g^{\mu\nu}(\partial_t\partial_\mu\partial_\nu\phi-\Gamma^\sigma_{\mu\nu}\partial_t\partial_\sigma \phi),
\end{eqnarray*}
because the metric (and hence the Christoffel symbols) are time-independent, due to the stationary symmetry. On the other side, we have
\begin{eqnarray*}
\Box\partial_t\phi=D_\mu g^{\mu\nu}\partial_\nu(\partial_t\phi)=g^{\mu\nu}(\partial_\mu\partial_\nu\partial_t \phi-\Gamma^\sigma_{\mu\nu}\partial_\sigma(\partial_t\phi)),
\end{eqnarray*}
which shows the result. Note that the action of a vector field on a function is still a function, i.e. $\mathcal L_K \phi=K^\mu\partial_\mu\phi$, and in the coordinate system in which $K^\mu=(1,0,0,0)$ this action reduces to $\mathcal L_K \phi=\partial_t \phi$ but the result remains a function and not a vector field.

\item The second property of the Killing vector field $K=\partial_\mu$ is antihermiticity.
Indeed,
\begin{align*}
(f,Kg)&=-i\int_\Sigma d^3x~~\sqrt{\gamma}~n^\mu(fD_\mu(\partial_tg^*)-(\partial_tg^*)D_\mu f),\\
&=-i\int_\Sigma d^3x~~\sqrt{\gamma}~n^\mu((-\partial_tf)D_\mu g^*-g^*D_\mu (-\partial_tf)),\\
&=(-Kf,g),
\end{align*}
where integration by parts was used in the second line. Since the operator $K$ is antihermitian, its eigenvalues are purely imaginary, i.e. $Kf_j=-i\omega f_j$, for some real $\omega$.
\end{enumerate}
This property of having purely imaginary eigenvalues is used to define the notion of positive frequency modes in the case of curved spacetime.
If
\begin{eqnarray}
\mathcal L_K f_j\equiv\partial_t f_j=-i\omega f_j,\quad \omega>0,
\end{eqnarray}
then the modes $\{f_j\}$ are called positive frequency modes. In the same way for modes $\{f_j^*\}$, i.e. if
\begin{eqnarray}
\mathcal L_K f_j^*\equiv\partial_t f_j^*=i\omega f_j^*,\quad \omega>0,
\end{eqnarray}
then the modes $\{f_j^*\}$ are called negative frequency modes.

The theory can be quantized, exactly in the same way as in flat space, namely by replacing classical fields by operators acting on Hilbert space, and by imposing canonical commutation relations between the operators.
Any field configuration $\phi(x)$ that is solution to the Klein-Gordon equation can be expanded with respect to the basis
\begin{eqnarray}
\phi(x)=\sum_i a_if_i+a_i^\dagger f_i^*.
\end{eqnarray}
To make things as simple as possible, the discrete notation is chosen. Again, the operators in front of positive and negative frequency modes in the field expansion are the creation/annihilation operators satisfying the commutation relation $[a_k,a_{k^\prime}^\dagger]=\delta(k-k^\prime)$. The vacuum state is defined to be the state $|0\rangle$ such that $a_k|0\rangle=0,~~\forall k$. In conclusion, stationary symmetry alows to pick up a preferred time coordinate, given by the timelike Killing vector field.

%\subsubsection{Non-stationary spacetime}
%--------------------------------------------------
%Positive frequency modes were defined because of the presence of a symmetry: the spacetime $(M,g)$ was assumed to be stationary, from which a Fock space was constructed. But one can wonder if it is still possible to construct a Fock space if the spacetime $(M,g)$ has no symmetry, but being still globally hyperbolic. One can show \cite{Carroll},\cite{Wipf} that a Fock space can be defined in this case as well if the two points function satisfies the so-called Hadamard condition,
%\begin{eqnarray}
%\langle\phi(x),\phi(y)\rangle=\frac{f_1}{\sigma}+f_2\ln \sigma+f_3,\notag
%\end{eqnarray}
%where $f_1,f_2,f_3$ are some smooth functions and where $\sigma$ is the square of geodesic distance between  points $x$ and $y$, i.e. $\sigma^2=g_{\mu\nu}(x^\mu-y^\mu)(x^\nu-y^\nu)$. Roughly speaking the Hadamard condition gives the form of the structure of the divergences of the two points function of the theory. If this condition is satisfied, then the Fock space can be build, even though the spacetime is not stationary.

This is the end of the general formalism of quantum field theory in curved spacetime. In the rest of the lecture, applications of it will be considered.

\subsection{Sandwich spacetime}
\begin{figure}[ht]
       \centering
      \includegraphics[width=5cm]{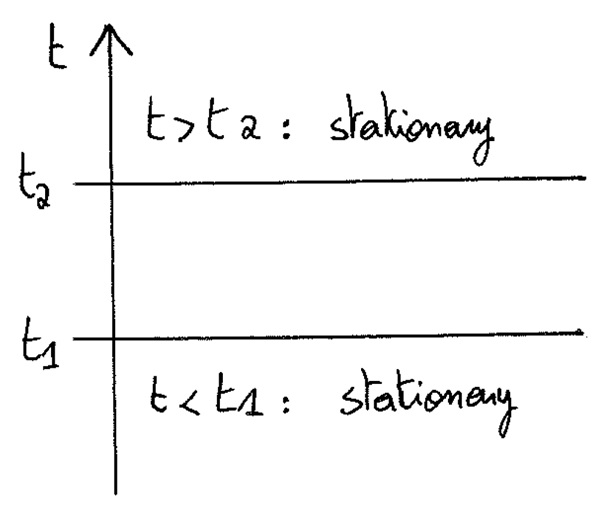}
     \caption{Sandwich spacetime: the spacetime is asymptotically stationary in past and future.}
    \label{fig8}
\end{figure}

%---------------------------------------
In this first application \cite{Townsend} of the formalism of the previous section, let us consider a spacetime $(M,g)$ composed by three regions called past, present and future. Furthermore the spacetime $(M,g)$ is supposed to be stationary in region past ($t<t_1$ in figure \ref{fig8}) and region future\footnote{This situation is somehow analogous to quantum field theory when studying the interaction between two fields \cite{Barnich}. One can suppose that the theory is free asymptotically in the past and asymptotically in the future when the fields do not interact.} ($t>t_2$ in figure \ref{fig8}). Let us finally suppose that the Klein-Gordon equation of motion hold throughout the space.

In region past, by hypothesis there exists a Killing vector field $K^P$  and by the results of the previous section positive frequency modes can be defined using this Killing vector, $\mathcal L_{K^P}f_j=-i\omega f_j, \omega>0$ such that they form an orthonormal basis $\{f_j,f_j^*\}$ and finally canonical quantization can be done. Similarly for region future, where the Killing vector field $K^F$ is used to defined positive frequency modes,  $\mathcal L_{K^P}g_j=-i\omega g_j, \omega>0$, forming a basis $\{g_j,g_j^*\}$ then canonical quantization can be achieved. Note that these two sets of modes, $\{f_j,f_j^*\}$ and $\{g_j,g_j^*\}$, are defined using the Killing vector fields $K^P$ and $K^F$ in regions past and future, respectively, but then they can be extended throughout spacetime by analytic continuation.

Any field configuration $\phi(x)$ solution to the Klein-Gordon equation of motion can be expanded in terms of the two bases,
\begin{eqnarray}\label{fieldexp}
\phi(x)=\sum_i(a_if_i+a_i^\dagger f_i^*)=\sum_i(b_ig_i+b_i^\dagger g_i^*),
\end{eqnarray}
where basis modes are normalized with respect to the Klein-Gordon inner product, and where operators $a_k,b_k$ in the expansion satisfy $[a_k,a_{k^\prime}^\dagger]=\delta(k-k^\prime)$ and $[b_p,b_{p^\prime}^\dagger]=\delta(p-p^\prime)$.

The modes $g_i$ can be expressed in terms of the basis $\{f_j,f_j^*\}$,
\begin{eqnarray}\label{bogo}
g_i=\sum_j (A_{ij}f_j+B_{ij}f_j^*)\label{2.35}.
\end{eqnarray}
This relation between the two bases is called a Bogoliubov transformation, and the coefficients $A,B$ inside the transformation are called the Bogoliubov coefficients.
Relations \eqref{2.35} imply
\begin{eqnarray}
g_i^*=\sum_j (B_{ij}^*f_j+A_{ij}^*f_j^*)\label{2.36}.
\end{eqnarray}
These Bogoliubov transformations can be written in a matrix form,
\begin{eqnarray}\label{bogomatrix}
\begin{pmatrix}
g\\g^*
\end{pmatrix}
=
\begin{pmatrix}
A&B\\B^*&A^*
\end{pmatrix}
\begin{pmatrix}
f\\f^*
\end{pmatrix}.
\end{eqnarray}

\subsubsection{Bogoliubov gymnastics}
%-----------------------------------------------
In order to invert relation \eqref{bogomatrix} easily, some relations between the Bogoliubov coefficients are needed.
Let us make some gymnastics with them.\\
The basis is normalized in such a way that $(\alpha f,\beta g)=\alpha\beta^*(f,g)$ see \eqref{2.20}, so we have
\begin{align*}
(g_i,g_j)&=\delta_{ij}\\
&=(A_{ip}f_p+B_{ip}f^*_p,A_{jq}f_q+B_{jq}f^*_q)\\
&=A_{ip}A_{jp}^*+B_{iq}B_{jq}^*(-1)=A_{ip}A_{pj}^\dagger-B_{ip}B_{pj}^\dagger.
\end{align*}
Hence we find the relation
\begin{eqnarray}
\fbox{$
AA^\dagger-BB^\dagger=1.\label{bogo1}
$}
\end{eqnarray}
From the basis orthonormalization, we also have
\begin{align*}
(g_i,g_j^*)&=0\\
&=(A_{ip}f_p+B_{ip}f_p^*,B_{jq}^*f_q+A_{jq}^*f_q^*)\\
&=A_{ip}B_{jp}+B_{iq}A_{jq}(-1)=A_{ip}B_{pj}^t-B_{ip}A_{pj}^t.
\end{align*}
Thus,
\begin{eqnarray}
\fbox{$
AB^t-BA^t=0.\label{bogo2}
$}
\end{eqnarray}
Relation \eqref{bogo1}\eqref{bogo2} between Bogoliubov coefficients allow to invert the matrix $M=\begin{pmatrix}A&B\\B^*&A^*\end{pmatrix}$ present in \eqref{bogomatrix} easily.
The inverse is given by
\begin{eqnarray}\label{inv}
M^{-1}=
\begin{pmatrix}A&B\\B^*&A^*\end{pmatrix}^{-1}=
\begin{pmatrix}A^\dagger&-B^t\\-B^\dagger&A^t\end{pmatrix}.
\end{eqnarray}
Indeed, we have
\begin{eqnarray*}
M^{-1}M=
\begin{pmatrix}
AA^\dagger-BB^\dagger&-AB^t+BA^t\\
B^*A^\dagger-A^*B^\dagger&-B^*B^t+A^*A^t
\end{pmatrix}
=\begin{pmatrix}
1&0\\
0&1
\end{pmatrix}.
\end{eqnarray*}
The field expansion \eqref{fieldexp} can be written in a matrix form,
\begin{align}\label{32}
\phi=
\begin{pmatrix}
b&b^\dagger
\end{pmatrix}
\begin{pmatrix}
g\\
g^*
\end{pmatrix}
=
\begin{pmatrix}
a&a^\dagger
\end{pmatrix}
\begin{pmatrix}
f\\
f^*
\end{pmatrix},
\end{align}
and using the Bogoliubov transformations \eqref{bogo} and \eqref{inv} we have
\begin{eqnarray}
\begin{pmatrix}
g\\g^*
\end{pmatrix}
=
\begin{pmatrix}
A&B\\B^*&A^*
\end{pmatrix}
\begin{pmatrix}
f\\
f^*
\end{pmatrix},\hspace{1cm}
\begin{pmatrix}
f\\f^*
\end{pmatrix}
=
\begin{pmatrix}
A^\dagger&-B^t\\-B^\dagger&A^t
\end{pmatrix}
\begin{pmatrix}
g\\
g^*
\end{pmatrix}.
\end{eqnarray}
So the field expansion \eqref{32} becomes
\begin{eqnarray}
\begin{pmatrix}
b&b^\dagger
\end{pmatrix}
\begin{pmatrix}
g\\
g^*
\end{pmatrix}
=
\begin{pmatrix}
a&a^\dagger
\end{pmatrix}
\begin{pmatrix}
A^\dagger&-B^t\\-B^\dagger&A^t
\end{pmatrix}
\begin{pmatrix}
g\\
g^*
\end{pmatrix},
\end{eqnarray}
which gives the relation
\begin{eqnarray}
\fbox{$
\begin{pmatrix}
b\\
b^\dagger
\end{pmatrix}
=
\begin{pmatrix}
A^*&-B^*\\-B&A
\end{pmatrix}
\begin{pmatrix}
a\\a^\dagger
\end{pmatrix}.
$}
\end{eqnarray}
This relation between the creation/annihilation operators with respect to the two different bases ends the Bogoliubov gymnastics part. Let us go back to physics.

\subsubsection{Particle number}
%-------------------------------------
The vacuum associated to modes $\{f_i,f_i^*\}$, called the $in$ vacuum, $|in\rangle$, is defined such that $a_i|in\rangle=0, \forall i$. Now the following question can be considered: what is the expected number of particles of species  $i$ present in the state $|in\rangle$ when evaluated by stationary observer in region future? Let us compute this number.
\begin{align*}
N_i&=\langle in|^FN_i|in\rangle=\langle in|b^\dagger_ib_i|in\rangle,\\
&=\langle in|(-B_{ip}a_p+A_{ip}a_p^\dagger)(A_{iq}^*a_q-B_{iq}^*a_q^\dagger)|in\rangle,\\
&=\langle in|B_{iq}B_{ip}^*\delta_{pq}|in\rangle=B_{ip}B_{pi}^\dagger.
\end{align*}
The number of particles is given by
\begin{eqnarray}\label{2.38}
\fbox{$
N_i=(BB^\dagger)_{ii}
$},
\end{eqnarray}
where there is no $i$ summation. So at the end one can see that the number of particles is just given by the Bogoliubov coefficient. And, in general, this coefficient is non-zero\footnote{This number is always zero in the case of unitary Bogoliubov transformation \eqref{bogo} as can be seen from \eqref{bogo1}, i.e. $B=0\Rightarrow AA^\dagger=1$.}. The total number of particles of all species is obtained by summing over all the species, i.e. $N=\sum_i N_i=Tr(BB^\dagger).$
\newpage

\section{Quantum Field Theory in Rindler spacetime in 1+1 dimensions -- Unruh effect}
%----------------------------------------------------------------------------------------------------
%----------------------------------------------------------------------------------------------------

In this part, quantum field theory in Rindler space \cite{Carroll} is studied. Many aspects will be relevant for the discussion of Hawking radiation later on. Quantum field theory in Rindler space means quantization carried out by an accelerating observer in Minkowski space. This will lead to the Unruh effect\footnote{Historically Unruh effect was discovered after Hawking radiation in order to better understand it. But as this lecture is not a lecture about history of physics, Unruh effect will be considered first, before Hawking effect.}. To evoid all possible complications, the quantum field theory is considered as simple as it can be, without becoming completely trivial, i.e. we consider a massless scalar field in two dimensions. The action is
\begin{align}
S=\int d^2x~ \left(-\frac{1}{2}\partial_\mu\phi\partial^\mu\phi\right)
\end{align}
from which the equations of motion read $\delta S=0\Rightarrow \Box \phi=0$.
Before quantizing the theory, let us spend some time studying Minkowski spacetime in two dimensions from the point of view of an accelerating observer.

\subsection{Rindler spacetime in 1+1 dimensions}
\begin{figure}[ht]
       \centering
      \includegraphics[width=4cm]{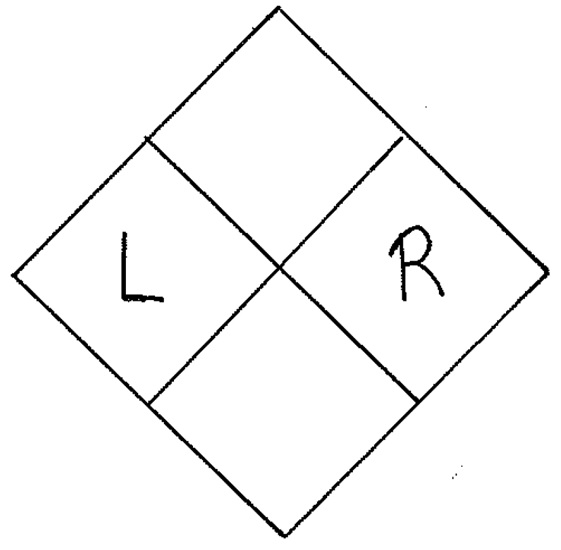}
     \caption{Left and right Rindler wedges in the Carter-Penrose diagram of Minkowski.}
    \label{fig7}
\end{figure}

%-----------------------------------------------------
By definition, Rindler spacetime is a sub-region of Minkowski spacetime ($ds^2=-dt^2+dx^2$) associated with an observer that is eternally accelerating at constant rate. The parametric motion (i.e. the trajectory) of such an observer is given by
\begin{align}
x(\tau)&=\frac{1}{\alpha}\cosh(\alpha\tau),\\
t(\tau)&=\frac{1}{\alpha}\sinh(\alpha\tau),
\end{align}
where $\alpha$ is a constant parameter. The acceleration is given by
\begin{align}
a^\mu&=\frac{D^2x^\mu}{d\tau^2}=\frac{d^2x^\mu}{d\tau^2}=(\alpha\sinh(\alpha\tau),\alpha\cosh(\alpha\tau)),\notag\\
a^2&=a^\mu a^\nu g_{\mu\nu}=\alpha^2,
\end{align}
and one can see that the acceleration is constant, $a=\pm \alpha$, as it should. The world line of the observer $x_\mu x^\mu$ satisfies $-t^2+x^2=\frac{1}{\alpha^2}$. This is an hyperbolic motion and lines $x=t$ are the horizons for this observer because region $x\leq t$ is forever forbidden to a Rindler observer.\\
In terms of the Carter-Penrose diagram of Minkowski space, the observer only covers two parts of it, called the Rindler wedges, the left one and right one, see figure \ref{fig7}.
\\

Instead of using coordinates $(t,x)$, let us introduce some new coordinates $(\eta,\xi)$ more fitted to the description of the accelerated motion in the right Rindler wedge,
\begin{align}\label{def}
t&=\frac{1}{a}e^{a\xi}\sinh(a\eta),\hspace{1cm} x=\frac{1}{a}e^{a\xi}\cosh(a\eta),\\
\eta&=\frac{\alpha}{a}\tau,\hspace{1cm}\xi=\frac{1}{a}\ln\frac{\alpha}{a}.
\end{align}
In terms of these coordinates, the proper time $\tau$ is proportional to $\eta$ and the spatial coordinate $\xi$ is constant. Moreover, an observer with acceleration rate $a=\alpha$ moves along $\eta=\tau,\xi=0$. In coordinates $(\xi,\eta)$, the metric becomes
\begin{align}
dt&= e^{a\xi}\sinh(a\eta)d\xi+e^{a\xi}\cosh(a\eta)d\eta\notag,\\
dx&= e^{a\xi}\cosh(a\eta)d\xi+e^{a\xi}\sinh(a\eta)d\eta\notag,\\
\Rightarrow ds^2&=-dt^2+dx^2=e^{2a\xi}(-d\eta^2+d\xi^2).
\end{align}
This metric is independent of $\eta$ so $\partial_\eta$ is a Killing vector field in these coordinates. The signification of this vector field is the following. In coordinates $(t,x)$ the vector $\partial_\eta$ is
\begin{align}
\partial_\eta&=\frac{\partial t}{\partial \eta}\partial_t+\frac{\partial x}{\partial \eta}\partial_x,\notag\\
&=e^{a\xi}\cosh(a\eta)\partial_t+e^{a\xi}\sinh(a\eta)\partial_x,\notag\\
&=a(x\partial_t+t\partial_x).
\end{align}
This is nothing else but a boost in the x direction.\\

All these considerations about the coordinates $(\xi,\eta)$ were done in the case of the right Rindler wedge. The same can be done in the case of the left Rindler wedge, by defining coordinates $(\eta,\xi)$ with opposite sign,
\begin{eqnarray}
t&=-\frac{1}{a}e^{a\xi}\sinh(a\eta),\hspace{1cm} x=-\frac{1}{a}e^{a\xi}\cosh(a\eta).
\end{eqnarray}
Because of the sign, the tangent vector to the hyperbolea $\partial_\eta$ is future pointing but in the opposite direction with respect to the right Rindler wedge.
Coordinates $(\eta,\xi)$ cannot be used simultaneously in wedges right and left, because the range of these parameters are the same in each regions. But then, why do we use the same set of coordinates twice instead of introducing new coordinates? The reason is that the metric is valid in both wedges and so it is more comfortable to work with.\\
It is worth to stress again the fact that the vector field $\partial_\eta$ is a killing vector field in wedges left and right, but is future pointing in the right wedge while past pointing in the left one.\\

An eternally accelerating observer allows to define three different globally hyperbolic manifolds equipped with future pointing Killing vectors, namely Minkowski space with Killing field $\partial_t$, right Rindler wedge with Killing field $\partial_\eta$ and left Rindler wedge with future pointing Killing field $-\partial_\eta$. Let us proceed to canonical quantization of the scalar field in Minkowski space and then in the right Rindler wedge.

\subsection{Quantization}
%------------------------------
\begin{itemize}
\item In Minkowski spacetime, the equation of motion reads $\Box\phi=(-\partial^2_t+\partial^2_x)\phi=0$ and admits plane waves solutions,
\begin{eqnarray}
f_k=\frac{1}{\sqrt{4\pi\omega}}e^{ik_\mu x^\mu},\hspace{1cm}k^\mu=(\omega,k).
\end{eqnarray}
Positive frequency modes are defined with respect to Killing vector field $\partial_t$, i.e. by the condition $\partial_t f_k=-i\omega f_k, \omega>0$. After canonical quantization, any field configuration $\phi$ solution to the equation of motion can be expanded in terms of $\{f_k,f_k^*\}$,
\begin{eqnarray}
\phi=\sum_k(a_kf_k+a_k^\dagger f_k^*).
\end{eqnarray}
The vacuum is defined by state $|0\rangle$ such that $a_k|0\rangle=0,\forall k$.

\item In the right Rindler wedge, the equation of motion reads $\Box\phi=e^{-2a\xi}(-\partial^2_\eta+\partial^2_\xi)\phi=0$ and admits plane wave solutions,
\begin{eqnarray}
g_k^R=\frac{1}{\sqrt{4\pi\omega}}e^{ik_\mu x^\mu},\hspace{1cm}x^\mu=(\eta,\xi).
\end{eqnarray}
Positive frequency modes are defined with respect the the Killing vector field $\partial_\eta$, i.e. by the condition $\partial_\eta g_k^R=-i\omega g_k^R,\omega>0.$
After canonical quantization, any field $\phi$ solution to the field equation can be expanded in terms of $\{g_k^R,g_k^{R*}\}$,
\begin{eqnarray}
\phi=\sum_k(b_kg_k^R+b_k^\dagger b_k^{R*}),
\end{eqnarray}
and the vacuum is defined by state $|^R0\rangle$ such that $b_k|^R0\rangle=0,\forall k$.
\end{itemize}

\subsection{Unruh effect}
%-----------------------------
Now the physically interesting question arrives: what does an observer in the right Rindler wedge see in the Minkowski vacuum?

Some care must be taken because the Rindler modes $g_k^R$ are not defined on a Cauchy surface for the whole Minkowski space. These modes are only defined on a Cauchy surface in the case of a manifold being the right Rindler wedge, and so these modes are not complete with respect to Minkowski spacetime. This is not really a problem because global modes can be defined,
\begin{eqnarray}
g_k^R=
\left\{
\begin{array}{rcr}
\frac{1}{\sqrt{4\pi\omega}}e^{ik_\mu x^\mu}&\hspace{1cm}\text{in right Rindler wedge}\\
0&\hspace{1cm}\text{in left Rindler wedge}
\end{array}
\right.
\end{eqnarray}
These modes are defined on an entire Cauchy surface ($\eta=0$ for instance), but they are not complete. Positive frequency modes need to be introduced in the left wedge, and are defined by
\begin{eqnarray}
g_k^L=
\left\{
\begin{array}{rcr}
0\hspace{1cm}\text{in right Rindler wedge}\\
\frac{1}{\sqrt{4\pi\omega}}e^{ik_\mu x^\mu}\hspace{1cm}\text{in left Rindler wedge}\\
\end{array}
\right.
\end{eqnarray}
These modes are positive frequency modes with respect to Killing vector field $-\partial_\eta$.\\

Taken together, $\{g_k^R,g_k^{R*}\}$ and $\{g_k^L,g_k^{L*}\}$ form a complete set of modes for the Minkowski space, and thus there are two possible modes expansions for any field configuration solution to the equation of motion,
\begin{align}
\phi&=\sum_k(b_kg_k^R+c_kg_k^L+h.c.),\\
\phi&=\sum_k(a_kf_k+h.c.).
\end{align}
Now one can wonder how many right particules are expected to be seen in Minkowski vacuum, or equivalently what does the eternally accelerating observer see in the Minkowski vacuum while being in the right Rindler wedge. Mathematically the question is
\begin{eqnarray}
\langle 0_{Mink}|^RN_k|0_{Mink}\rangle=~~?\notag
\end{eqnarray}
There is a relation between modes $g_k^R$ and $f_k$, given by the Bogoliubov transformation\footnote{Here the Bogoliubov transformation is written in continuous notation, for convenience.},
\begin{eqnarray}\label{3.20}
g_k^R(u)=\int d\omega^\prime(A_{\omega\omega^\prime}f_{\omega^\prime}+B_{\omega\omega^\prime}f_{\omega^\prime}^*).
\end{eqnarray}
Recall that $f_{\omega^\prime}=\frac{1}{\sqrt{4\pi\omega^\prime}}e^{ik_\mu x^\mu}$ and $ik_\mu x^\mu=-i\omega^\prime t+ikx=-i\omega^\prime(t-x)=-i\omega^\prime u$ with $u=t-x$. So the Bogoliubov transformation relating the bases \eqref{3.20} becomes
\begin{eqnarray}
g_k^R(u)=\int d\omega^\prime\left(A_{\omega\omega^\prime}\frac{1}{2\pi}\sqrt{\frac{\pi}{\omega^\prime}}e^{-i\omega^\prime u}+B_{\omega\omega^\prime}\frac{1}{2\pi}\sqrt{\frac{\pi}{\omega^\prime}}e^{i\omega u}\right).
\end{eqnarray}
This expression for $g_k^R(u)$ looks like the inverse Fourier transform of $g_k^R(u)$. Indeed,
\begin{eqnarray}\label{3.22}
g_k^R(u)=\frac{1}{2\pi}\int_{-\infty}^{+\infty}d\omega^\prime e^{-i\omega^\prime u}\tilde g_\omega(\omega^\prime),
\end{eqnarray}
where $\tilde g_\omega(\omega^\prime)$ is the Fourier transform of $g_k^R(u)$, i.e. $\tilde g_\omega(\omega^\prime)=\int_{-\infty}^{+\infty}due^{i\omega^\prime u}g_k^R(u)$. Equation \eqref{3.22} can also be re-written as
\begin{eqnarray}\label{3.23}
g_k^R(u)=\frac{1}{2\pi}\int_{0}^{+\infty}d\omega^\prime e^{-i\omega^\prime u}\tilde g_\omega(\omega^\prime)+\frac{1}{2\pi}\int_{0}^{+\infty}d\omega^\prime e^{i\omega^\prime u}\tilde g_\omega(-\omega^\prime),
\end{eqnarray}
where the values of integration and also the variable of integration were changed in the second term. By comparing this expression \eqref{3.23}, i.e. the expression for the inverse Fourier transform of the function $g_k^R(u)$, with the expression \eqref{3.20}, i.e. the expression for the Bogoliubov transformation of the function $g_k^R(u)$, one can get an expression for the Bogoliubov coefficient,
\begin{eqnarray}\label{3.24}
A_{\omega\omega^\prime}=\sqrt{\frac{\omega^\prime}{\pi}}\tilde g_\omega(\omega^\prime),\hspace{1cm}B_{\omega\omega^\prime}=\sqrt{\frac{\omega^\prime}{\pi}}\tilde g_\omega(-\omega^\prime).
\end{eqnarray}
Recall the relation $AA^\dagger-BB^\dagger=1$ from the Bogoliubov gymnastics \eqref{bogo1}, which is equivalent to $|A|^2-|B|^2=1$. So if there also exists a relation between $\tilde g(-\omega^\prime)$ and $\tilde g(\omega^\prime)$ then the Bogoliubov coefficient $|B|^2$ will be immediately known without other computation.

The desired relation between functions $\tilde g(-\omega^\prime)$ and $\tilde g(\omega^\prime)$ is given by
\begin{eqnarray}\label{3.25}
\tilde g(-\omega^\prime)=-e^{-\pi\omega/a}\tilde g(\omega^\prime).
\end{eqnarray}

\subsubsection{Fun with integrals}
%----------------------------------------
In this section the relation \eqref{3.25} is proved,
\begin{eqnarray}
\tilde g(-\omega^\prime)=-e^{-\pi\omega/a}\tilde g(\omega^\prime).
\end{eqnarray}
The Fourier transform of $g_\omega^R(u)$ is
\begin{eqnarray}\label{3.26}
\tilde g_\omega(\omega^\prime)=\int_{-\infty}^{+\infty}du~ e^{i\omega^\prime u}g_\omega^R(u),
\end{eqnarray}
with $g_\omega^R(u)=\frac{1}{\sqrt{4\pi\omega}}e^{ik_\mu x^\mu}$ defined in the right Rindler wedge, i.e. for $u<0$. The right Rindler wedge is equipped with coordinates $(\eta,\xi)$, so that $ik_\mu x^\mu=-i\omega(\eta-\xi)$. Now a relation must be found between coordinates $(t,x)$ and the difference $(\eta-\xi)$. From the definition \eqref{def} we get
\begin{align}
t-x=\frac{1}{a}e^{a\xi}(-e^{-a\eta})\hspace{1cm}\Rightarrow\hspace{1cm}u&=-\frac{1}{a}e^{-a(\eta-\xi)}\notag\\
\eta-\xi&=-\frac{1}{a}\ln(-au).\label{3.27}
\end{align}
Taking \eqref{3.27} into account, the Fourier transform of $g_\omega^R(u)$, \eqref{3.26}, becomes
\begin{align}
\tilde g_\omega(\omega^\prime)&=\int_{-\infty}^{0}du~~ e^{i\omega^\prime u}\frac{1}{\sqrt{4\pi\omega}}~~e^{(i\omega/a)\ln(-au)},\notag\\
&=\frac{1}{\sqrt{4\pi\omega}}\int_{-\infty}^{0}du ~~e^{i\omega^\prime u}~~(-au)^{i\omega/a},\notag\\
&=\frac{1}{\sqrt{4\pi\omega}}a^{i\omega/a}\int^{+\infty}_{0}du~~e^{-i\omega^\prime u}u^{i\omega/a},\notag\\
&=\frac{1}{\sqrt{4\pi\omega}}a^{i\omega/a}\frac{1}{a}\frac{\omega}{\omega^\prime}\int_0^{+\infty}e^{-i\omega^\prime u}u^{(i\omega/a)-1}.\label{L4}
\end{align}
Where a change of variable occurred in the third line and an integration by parts took place in \eqref{L4}. Before continuing further with $g_\omega^R(u)$, let us consider for a while the following integral  \cite{Mukhanov}
\begin{eqnarray}\label{3.32}
\int_0^\infty dx ~~e^{-bx}x^{s-1},
\end{eqnarray}
defined for positive real part of parameters $b$ and $s$. This integral \eqref{3.32} can be rewritten as
\begin{eqnarray}\label{3.33}
\int_0^\infty \frac{d(bx)}{b}e^{-bx}\frac{(bx)^{s-1}}{b^{s-1}}=b^{-s}~~ \int_0^\infty dy~~e^{-y}y^{s-1}=e^{-s\ln b}\Gamma(s),
\end{eqnarray}
where the definition of the gamma function was used in the last equality, i.e. $\Gamma(s)=\int_0^\infty dy~~e^{-y}y^{s-1}$.
In this integral, the parameter $b$ is a complex number and consequently $\ln b$ is a multivalued function\footnote{Indeed: $\ln b=\ln(r{e^{i(\theta+k\pi)}})=\ln r+i(\theta+k\pi), \forall k\in \mathcal Z$.}. The definition adopted here for this function is the following. For a complex number $b$ there exists some cartesian reals numbers $(A,B)$ and corresponding polar coordinates $(r,\theta)$ such that $b=A+iB=re^{i\theta}$. Therefore $\ln b$ is defined to be
\begin{align}
\ln b&=\ln(A+iB)=\ln(re^{i\theta})=\ln r+i\theta=\ln\sqrt{A^2+B^2}+i\arctan \left(\frac{B}{A}\right),\notag\\
&=\ln\sqrt{A^2+B^2}+i\arctan \left|\frac{B}{A}\right|~~sign\left(\frac{B}{A}\right).\label{3.34}
\end{align}
This is the definition adopted for $\ln b$ with a complex number $b$.\\

This result for the integral \eqref{3.32} can be used to solve the initial integral \eqref{L4}, with parameters $b$ and $s$ given by $b=i\omega^\prime$ and $s=i\omega/a$.  But in order to use the result \eqref{3.33}, the real part of parameters $b$ and $s$ must be positive. This condition is satisfied by introducing a small real positive parameter $\epsilon$,
\begin{eqnarray}\label{3.35}
b=i\omega^\prime+\epsilon,\hspace{1cm}s=i\omega/a+\epsilon,
\end{eqnarray}
and then taking $\lim_{\epsilon\to 0^+}$. Now $\ln b$ can be computed, using \eqref{3.34} with $b$ given by \eqref{3.35},
\begin{align}
\ln b&=\ln\sqrt{(\omega^\prime)^2+\epsilon^2}+i\arctan\left|\frac{\omega^\prime}{\epsilon}\right| sign\left(\frac{\omega^\prime}{\epsilon}\right)\notag\\
\lim_{\epsilon\to 0^+} b&=\ln|\omega^\prime|+i\frac{\pi}{2} sign(\omega^\prime).
\end{align}
The final result for the Fourier transform of $g_\omega^R(u)$, valid for all $\omega^\prime$, is thus
\begin{eqnarray}\label{3.38}
\tilde g_\omega(\omega^\prime)=\frac{1}{\sqrt{4\pi\omega}}a^{i\omega/a}\left(\frac{1}{a}\frac{\omega}{|\omega^\prime|}sign(\omega^\prime)\right)e^{-i\omega/a\ln |\omega^\prime|}~~e^{\omega\pi/(2a)sign(\omega^\prime)}~~\Gamma\left(\frac{i\omega}{a}\right).
\end{eqnarray}
For $\omega^\prime>0$ the two different functions $\tilde g_\omega(\omega^\prime)$ and $\tilde g_\omega(-\omega^\prime)$ can be computed using the result of \eqref{3.38},
\begin{align}
\tilde g_\omega(\omega^\prime)&=\frac{1}{\sqrt{4\pi\omega}}a^{i\omega/a}\left(\frac{1}{a}\frac{\omega}{|\omega^\prime|}\right)e^{-i\omega/a\ln |\omega^\prime|}~~e^{\omega\pi/(2a)}~~\Gamma\left(\frac{i\omega}{a}\right),\label{3.39}\\
\tilde g_\omega(-\omega^\prime)&=\frac{1}{\sqrt{4\pi\omega}}a^{i\omega/a}\left(-\frac{1}{a}\frac{\omega}{|\omega^\prime|}\right)e^{-i\omega/a\ln |\omega^\prime|}~~e^{-\omega\pi/(2a)}~~\Gamma\left(\frac{i\omega}{a}\right).\label{3.40}
\end{align}
Comparison between \eqref{3.39} and \eqref{3.40} implies the result
\begin{eqnarray}
\tilde g_\omega(-\omega^\prime)=-e^{-\omega\pi/a}\tilde g_\omega(\omega^\prime),\notag
\end{eqnarray}
which is exactly the relation \eqref{3.25}.

\subsubsection{Thermal bath}
%----------------------------------------
The Bogoliubov coefficients \eqref{3.24}, together with the relation \eqref{3.25}, yield a simple relation between them,
\begin{eqnarray}\label{3.43}
A_{\omega\omega^\prime}=\sqrt{\frac{\omega^\prime}{\pi}}\tilde g_\omega(\omega^\prime)=-\sqrt{\frac{\omega^\prime}{\pi}}\tilde g_\omega(-\omega^\prime)e^{\pi\omega/a}=-e^{\pi\omega/a}B_{\omega\omega\prime}.
\end{eqnarray}
This result, with the relation $|A|^2-|B|^2=1$ derived in \eqref{bogo1}, implies
\begin{eqnarray}\label{456}
|B|^2=\frac{1}{e^{2\pi\omega/a}-1}.
\end{eqnarray}
This Bogoliubov coefficient is exactly the number of particles seen by a right Rindler observer in the Minkowski vacuum, cfr. equation \eqref{2.38} in the general case. Relation \eqref{456} is a Planck spectrum with temperature $T=a/2\pi$. This result shows that a Rindler observer is immersed in a thermal bath of particles, and it is called the Unruh effect.

\section{Quantum Field Theory in spacetime of spherically collapsing star -- Hawking effect}
%---------------------------------------------------------------------------------------------------------------
%---------------------------------------------------------------------------------------------------------------
\subsection{The set-up}
\begin{figure}[ht]
       \centering
      \includegraphics[width=6cm]{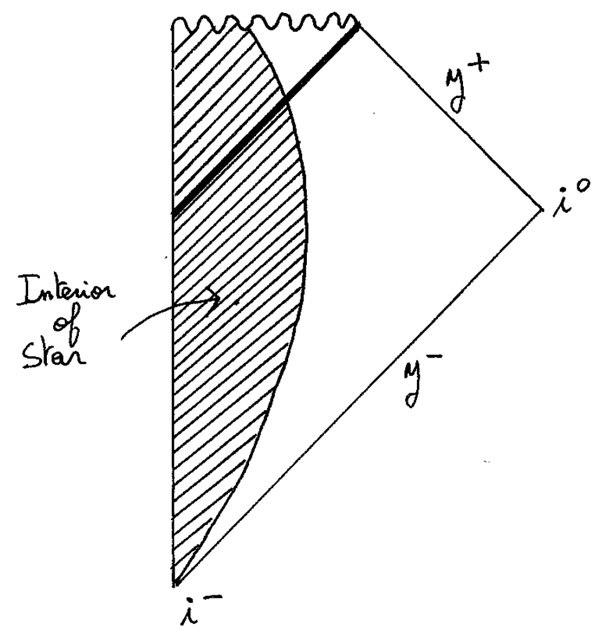}
     \caption{Carter-Penrose diagram of a spherically collapsing star.}
    \label{fig9}
\end{figure}

%---------------------------
In this section we consider a spacetime corresponding to a spherically collapsing star, following the original article by Hawking \cite{HawkingP} and also \cite{Dowker}. The CP diagram is given by figure \ref{fig9}.

Even though Schwarzschild spacetime is static, the collapsing star is not because it involves complicated internal dynamics. However the spacetime of a collapsing star is stationary in past null infinity ($\mathcal J^-$) and in future null infinity ($\mathcal J^+$). So the situation is exactly the same as in the sandwich spacetime, considered in the first chapter.

A massless scalar field propagates throughout the spacetime and canonical quantization can be performed with respect to a basis of modes defined at $\mathcal J^-$ and also with respect to a set of modes defined at $\mathcal J^+$, exactly as in the case of the sandwich spacetime. Then a physically interesting question we can ask is: does an observer in far future see particles in the vacuum defined in the past?

From the CP diagram, one can see that $\mathcal J^-$ is a Cauchy surface but $\mathcal J^+$ is not, because of the region behind the horizon. However $\mathcal J^+ \bigcup \mathcal H^+$ (with $\mathcal H^+$ the future horizon\footnote{There is no timelike Killing on $\mathcal H^+$, so the notion of positive frequency modes cannot be defined in the usual way. But computations of this section do not depend on this fact.}) is a Cauchy surface for the spacetime.

So three sets of modes can be defined,
\begin{itemize}
\item $\{f_i\}$, positive frequency modes on $\mathcal J^-$,
\item $\{g_i\}$, positive frequency modes on $\mathcal J^+$ and no Cauchy datum on $\mathcal H^+$,
\item $\{h_i\}$, positive frequency modes on $\mathcal H^+$ and no Cauchy datum on $\mathcal J^+$.
\end{itemize}
Modes $\{f_i,f_i^*\}$ form a complete set, and $\{g_i,g_i^*\}\bigcup\{h_i,h_i^*\}$, when taken together also form  a complete set.

The next stage is to expand any field configuration $\phi$ solution to the equation of motion in terms of the two different complete sets of modes,
\begin{eqnarray}
\phi=\sum_i (a_if_i+h.c.)=\sum_i(b_ig_i+c_ih_i+h.c.).
\end{eqnarray}
The vacuum with respect to the basis $\{f_i,f_i^*\}$ is defined to be the state $|in\rangle$ such that $a_i|in\rangle=0,\forall i$, and the number of particles seen by an observer in far future in the $|in\rangle$ vacuum is given by $N_i=(BB^\dagger)_{ii}$. The Bogoliubov coefficient of the expansion $g_i=\sum_j(A_{ij}f_j+B_{ij}f^*_j)$ is thus needed in order to compute the particles number we are looking for.
A natural way of computing this Bogoliubov coefficient is first to find a solution to the equation of motion in the Schwarzschild background and then to go to Fourier space doing the same analysis as in the case of the Rindler space.

The problem of this straightforward approach is that there is no easy solution to the equation of motion in the Schwarzschild background, as will be seen in the next section.

\subsection{Wave equation}
%--------------------------------
The schwarzschild metric is
\begin{eqnarray}
ds^2=-\left(1-\frac{2M}{r}\right)dt^2+\left(1-\frac{2M}{r}\right)^{-1}dr^2+r^2d\Omega^2,
\end{eqnarray}
with $d\Omega^2=d\theta^2+\sin^2\theta d\phi^2$. We have $\sqrt{-g}=r^2\sin\theta$ and the wave equation $\Box \varphi =0$ reads
\begin{align}
\Box\varphi&=\frac{1}{\sqrt{-g}}\partial_\mu\left(g^{\mu\nu}\sqrt{-g}\partial_\nu\varphi\right),\notag\\
&=\partial_t\left(-\left(1-\frac{2M}{r}\right)^{-1}\partial_t\varphi\right)+\frac{1}{r^2}\partial_r\left(\left(1-\frac{2M}{r}\right)r^2\partial_r\varphi\right)+\frac{1}{r^2}\Box_{S^2}\varphi,\label{4.4}
\end{align}
with $\Box_{S^2}\varphi=\left(\frac{1}{\sin\theta}\partial_\theta(\sin\theta\partial_\theta)+\frac{1}{\sin^2\theta}\partial_\phi^2\right)\varphi$.
The ansatz $\varphi=\frac{f(r,t)}{r}Y^m_l(\theta,\phi)$ reduces the wave equation \eqref{4.4} to
\begin{eqnarray}
 -\left(1-\frac{2M}{r}\right)^{-1}\partial_t^2f+\frac{2M}{r^2}\left(\partial_r f-\frac{1}{r}f\right)+\left(1-\frac{2M}{r}\right)\partial_r^2f-\frac{l(l+1)}{r^2}f=0.\label{4.5}
\end{eqnarray}
The last term comes from the property of the spherical harmonics, $\Box_{S^2} Y^m_l=-l(l+1)Y_l^m$.
Let us introduce the tortoise coordinate $r^*$ by
\begin{align}
dr^*&=\frac{dr}{1-\frac{2M}{r}}\label{4.6}\\
\partial_r&=\frac{\partial_{r^*}}{\partial_r}\partial_{r*}=\left(1-\frac{2M}{r}\right)^{-1}\partial_{r*}\notag\\
\partial_r^2&=-\left(1-\frac{2M}{r}\right)^{-2}\frac{2M}{r^2}\partial_{r^*}+\left(1-\frac{2M}{r}\right)^{-2}\partial_{r^*}^2.\notag
\end{align}
In terms of the tortoise coordinate the wave equation \eqref{4.5} takes the form
\begin{eqnarray}
(-\partial_t^2+\partial_{r^*}^2)f-\left(1-\frac{2M}{r}\right)\left(\frac{l(l+1)}{r^2}+\frac{2M}{r^3}\right)f=0.
\end{eqnarray}
Let us call the potential $V(r)=\left(1-\frac{2M}{r}\right)\left(\frac{l(l+1)}{r^2}+\frac{2M}{r^3}\right)$. Then the wave equation becomes
\begin{eqnarray}
(-\partial_t^2+\partial_{r^*}^2)f=V(r)f.
\end{eqnarray}
The tortoise coordinate $r^*$ defined by \eqref{4.6} is given explicitly in terms of the coordinate $r$ by
\begin{eqnarray}
r^*=\int dr\frac{r-2M+2M}{r-2M}=r+2M\ln(r-2M)+C=r+2M\ln\left(\frac{r}{2M}-1\right),\notag
\end{eqnarray}
where the integration constant $C$ was chosen to be $C=-2M\ln (2M)$. The name tortoise comes from the derivative
\begin{eqnarray}
\frac{dr}{dr^*}=1-\frac{2M}{r},\hspace{1cm}\lim_{r\to 2M}\frac{dr}{dr^*}=0,\notag
\end{eqnarray}
which means that the function $r=r(r^*)$ becomes more and more constant as one approaches the horizon, hence the name.

Let us look at the form of the potential $V$,
\begin{itemize}
\item $r\to 2M ~~(\Leftrightarrow r^*=-\infty): V=0,$
\item $r\to \infty ~~(\Leftrightarrow r^*=+\infty): V=0$,
\end{itemize}
and between these two values of $r$, the precise form of the potential $V$ depends on $l$, and represents a potential barrier, so that any solution coming from infinity is expected to be partially reflected and partially emitted. At $\mathcal J^+$ the potential is zero and the wave equation reduces to
\begin{eqnarray}\label{eq3.8}
(-\partial^2_t+\partial_{r^*}^2)f=0,
\end{eqnarray}
 whose solutions are plane waves\footnote{Note that plane waves are delocalized (i.e. they have a support everywhere on $\mathcal J^+$) but wave packets (i.e. a linear combination of planes waves)  can be constructed on $\mathcal J^+$ around some momentum $\omega_0$ \cite{Dowker} by using the superposition principle .}.

Later on, outgoing plane wave solutions to \eqref{eq3.8} will be considered,
\begin{align}
f&=e^{ik_\mu x^\mu},\hspace{1cm}k_\mu=(-\omega,k),x^\mu=(t,r^*)\notag\\
&=e^{-i\omega u}\label{4.15}
\end{align}
Now an approximation\footnote{Cfr. the original paper by Hawking \cite{HawkingP}.} is made, the geometric optics approximation.

\subsection{Geometric optics approximation}
%-----------------------------------------------------
A general wave $f$ given by $f=a e^{iS}$ is such that $a$ is constant with respect to the phase $S$, this is the geometric optics approximation \cite{Landau}. The wave equation $\Box f=0$ implies
\begin{align}
0=D_\mu D^\mu (ae^{iS})&=D_\mu(i(D^\mu S) f)=i(D_\mu D^\mu S) f-(D_\mu S)(D^\mu S) f\notag\\
&=iSD_\mu D^\mu f-(D_\mu S)(D^\mu S) f,\label{4.17}
\end{align}
where double integration by parts were done to get the first term of \eqref{4.17}. But then the equation of motion $\Box f=0$ implies that  \eqref{4.17} reduces to
\begin{eqnarray}\label{4.18}
D_\mu S D^\mu S=0.
\end{eqnarray}
Suppose moreover that $S(x)$ is a family of surfaces of constant phase. The normal vector to $S(x)$ is $k_\mu=\partial_\mu S$ and condition \eqref{4.18} is equivalent to the condition that the normal vector is null,
\begin{eqnarray}\label{4.19}
k_\mu k^\mu=0.
\end{eqnarray}
Due to its light-like nature, the vector $k_\mu$ is also normal to some tangent vector $k^\mu$ for a certain curve $x^\mu(\lambda)$ lying in the surface $S(x)$, i.e. $k^\mu=\frac{dx^\mu}{d\lambda}$.

A special property of $k^\mu$ is that it is the tangent vector of a very specific class of curves, namely $k^\mu$ is the tangent vector field of null geodesic curves.
Indeed, by taking the covariant derivative of \eqref{4.19}
\begin{align}
0=k_\mu k^\mu\hspace{1cm}\Rightarrow\hspace{1cm}0&=2k^\mu D_\sigma k_\mu=2(k^\mu D_\sigma(\partial_\mu S))=2(k^\mu D_\mu(\partial_\sigma S))\notag\\
0&=k^\mu D_\mu k_\sigma.\label{4.22}
\end{align}
Equation \eqref{4.22} is the geodesic equation. Indeed,
\begin{eqnarray}
0=k^\mu D_\mu k^\sigma=k^\mu \partial_\mu k^\sigma+k^\mu  \Gamma^\sigma_{\mu\nu}k^\nu=\frac{d^2x^\sigma}{d\lambda^2}+\Gamma^\sigma_{\mu\nu}\frac{dx^\mu}{d\lambda}\frac{dx^\nu}{d\lambda},\notag
\end{eqnarray}
where the fact that $k^\mu$ is the tangent vector to a certain curve was used, $k^\mu=\frac{dx^\mu}{d\lambda}$.

In conclusion, the geometric optics approximation implies that the surface of constant phase of any wave $f=ae^{iS}$ can be traced back in time by following null geodesics. That is the approximation that Hawking did in his original paper, and that is the approximation that is done in the next section.

\subsection{Hawking's computation}
\begin{figure}[ht]
       \centering
      \includegraphics[width=8cm]{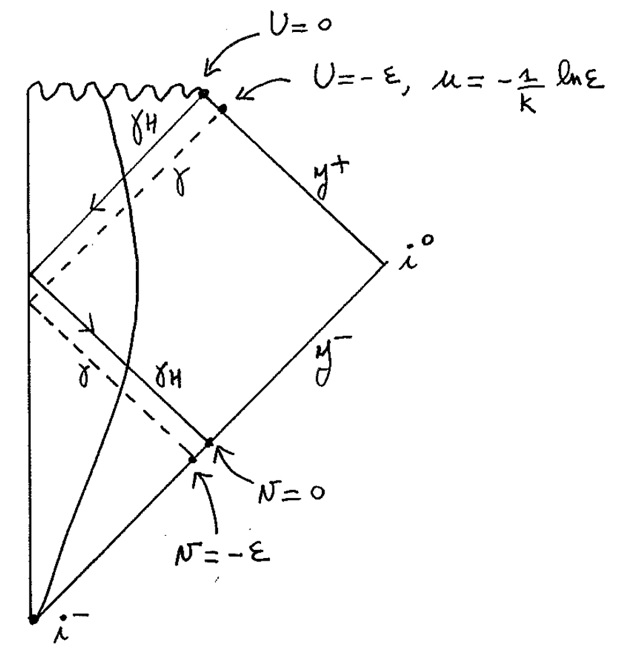}
     \caption{Outgoing solution traced back in time by following null geodesic $\gamma$ from $\mathcal J^+$ to $\mathcal J^-$.}
    \label{fig10}
\end{figure}

%------------------------------------------
At $\mathcal J^+$ an outgoing solution to the wave equation was found to be $f=e^{-i\omega u}$, cfr. \eqref{4.15}. The geometric optics approximation is used\footnote{This approximation will be justified later.} to trace back in time the outgoing solution by following null geodesics, see figure \ref{fig10}.

\begin{figure}[ht]
       \centering
      \includegraphics[width=6cm]{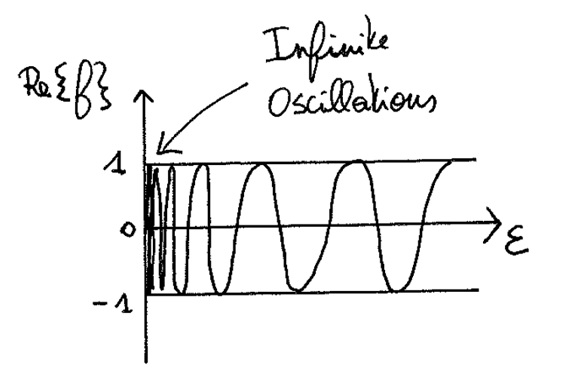}
     \caption{Justification of the geometric optics approximation.}
    \label{fig11}
\end{figure}

Let $\gamma_H$ denote the limiting null geodesic staying at the horizon, and suppose that the null geodesic $\gamma$ associated with the outgoing solution \eqref{4.15} is close to $\gamma_H$. The affine distance at $\mathcal J^+$ is $U$ (the Kruskal coordinate), and is related to $u=t-r^*$ by the relation
\begin{eqnarray}\label{4.24}
U=-e^{-u\kappa},
\end{eqnarray}
 with $\kappa$ the surface gravity (cfr. section 5 for further details about this point). At $\mathcal J^+$, if the affine distance of the limiting null geodesic $\gamma_H$ is chosen to be $U=0$, then the one of the null geodesic $\gamma$ of the outgoing solution is $U=-\epsilon$, see figure \ref{fig10}. From the relation \eqref{4.24} the value of $u$ for the geodesic $\gamma$ can be deduced to be
\begin{eqnarray}\label{rel}
u=-\frac{1}{\kappa}\ln\epsilon.
\end{eqnarray}
This coordinate $u$ is the affine distance between $\gamma_H$ and $\gamma$ along an ingoing null geodesic. The outgoing plane wave that is considered at $\mathcal J^+$ can thus be rewritten as
\begin{eqnarray}\label{4.26}
f=e^{-i \omega u}=e^{i\omega/\kappa \ln\epsilon}.
\end{eqnarray}
Near the horizon (i.e. $\gamma\to\gamma_H$ which means $\epsilon\to 0$) the geometric optics approximation is valid, since there is an infinite oscillation in the phase of \eqref{4.26}, see figure \ref{fig11}.

According to the geometric optics approximation, surfaces of constant phase of the solution \eqref{4.26} can be traced back in time by following the null geodesic $\gamma$. When tracing back in time, $\gamma$ will reach $\mathcal J^-$ at an affine distance $v=-\epsilon$ with respect to the limiting null geodesic $\gamma_H$. So
\begin{itemize}
\item at $\mathcal J^+$, the outgoing solution is $f=e^{i\omega/\kappa \ln\epsilon}$,
\item at $\mathcal J^-$, the outgoing solution is $f=e^{i\omega/\kappa \ln-v}$.
\end{itemize}
This solution at $\mathcal J^-$ has exactly the same form as the one for the Rindler space case, but now with $\kappa$ instead of $a$. The same business as for Rindler space can be done, i.e. going into Fourier space, comparing with Bogoliubov expansion and using funny integrals to conclude that the number of particles is given by
\begin{eqnarray}
|B|^2=\frac{1}{e^{2\pi\omega/\kappa}-1},\hspace{1cm}T=\kappa/2\pi.
\end{eqnarray}
This is a Planck spectrum with temperature $T=\kappa/2\pi$; this is the Hawking radiation: a black hole has a temperature and its thermal emission leads to a decrease in the mass of the black hole and possibly to its evaporation.

An important remark is that during this presentation, the potential barrier of the potential $V$ was not taken into account. This is the backreaction problem and when it is considered the spectrum is modified in the following way
\begin{eqnarray}
|B|^2=\frac{1}{e^{2\pi\omega/\kappa}-1}\Gamma_\omega,
\end{eqnarray}
where $\Gamma_\omega$ is the greybody factor, depending on the spin and angular part of the potential $V$. The important point to observe in this case is that the spectrum is no longer thermal when the backreaction is taken into account.
On the other side, if the backreaction problem is not considered then the precise form of the potential $V$ in $-\partial_t^2+\partial_{r^*}^2f=Vf$ is totally irrelevant in the geometric optics approximation.

In the next section some consequences of Hawking radiation are studied.

\newpage
\section{Some consequences of Hawking radiation}
%-------------------------------------------------------------
%------------------------------------------------------------- 
\subsection{Thermodynamics of black holes}
Before Hawking discovered that black holes radiate at temperature $T=\kappa/2\pi$, there was an analogy between the laws of thermodynamics and the laws of black holes. Then he proved that black holes are actually thermal objects, so the initial analogy is more than just an analogy.

The zeroth law of thermodynamics states that the temperature $T$ of a body in thermal equilibrium  is constant. One can wonder what is the analogous for the black hole but of course from the result presented in the previous section, the answer is already known. In this section $\kappa$ is shown to be constant on the black hole horizon.

\subsubsection{Surface gravity}
%-------------------------------------
In this part, let us consider \cite{Townsend} an hypersurface $S(x)$, the normal of which is $l^\mu=g^{\mu\nu}\partial_\nu S$ and is null, $l_\mu l^\mu=0$. From \eqref{4.22} it is already known that $l^\mu$ is the tangent vector field to null geodesics curves.\\

\noindent \textsc{Definition 7}\\
A null hypersurface $\Sigma$ is called Killing horizon of a Killing vector field $\xi$ if the Killing vector field $\xi$ is proportional to $l$, namely $\xi^\mu=fl^\mu$, for some smooth function $f$.\\

Let us compute $\xi^\mu D_\mu \xi^\sigma$, remembering the geodesic equation \eqref{4.22},
\begin{eqnarray}\label{5.1}
\xi^\mu D_\mu \xi^\sigma=(fl^\mu) (D_\mu f)l^\sigma+(fl^\mu) f D_\mu l^\sigma=(fl^\mu) (\partial_\mu f)l^\sigma=(\xi^\mu\partial_\mu\ln f)\xi^\sigma.
\end{eqnarray}
The surface gravity $\kappa$ is defined by $\kappa=\xi^\mu\partial_\mu\ln f$, then \eqref{5.1} becomes
\begin{eqnarray}\label{5.2}
\fbox{$
\xi^\mu D_\mu \xi^\sigma=\kappa \xi^\sigma
$}.
\end{eqnarray}
This relation allows to compute the surface gravity more easily than using its definition. For instance, for the Schwarzschild black hole in ingoing Eddington-Finkelstein coordinates\footnote{The computation is easier in this coordinates system.}, the metric reads
\begin{eqnarray}
ds^2=-\left(1-\frac{2M}{r}\right)dv^2+2dvdr+r^2d\Omega^2,\notag
 \end{eqnarray}
with $\partial_v$ a (timelike) Killing vector, $K=K^\mu\partial_\mu$ with $K^\mu=(1,0,0,0)$.  The computation of \eqref{5.2} with $\sigma=v$ gives an expression for the surface gravity,
\begin{eqnarray}
\xi^\sigma D_\sigma \xi^v=\xi^\sigma \partial_\sigma \xi^v+\xi^\sigma\Gamma_{\sigma\kappa}^v\xi^\kappa=\Gamma^v_{vv}=-\frac{1}{2}g^{vr}g_{vv,r}=\frac{M}{r^2}.\notag
\end{eqnarray}
This result should be evaluated at the horizon, by taking $r=2M$, and so the surface gravity for the Schwarzschild black hole is given by
\begin{eqnarray}
\fbox{$
\kappa=\frac{1}{4M}.
$}
\end{eqnarray}

\subsubsection{Some properties of $\kappa$}
%-----------------------------------------------------
One can wonder why $\kappa$ is called the surface gravity, this is the purpose of this sub-section. In order to do this, the fact that $\kappa$ is constant on the horizon is needed, requiring to establish firstly two other properties of $\kappa$.
\begin{enumerate}
\item First, the Killing vector of a Killing horizon is by definition orthogonal to a hypersurface. Frobenius theorem can then be used to say that
\begin{eqnarray}\label{5.6}
\xi_{[\mu} D_\nu \xi_{\sigma]}=0,
\end{eqnarray}
where brackets denote antisymmetrization. The fact that $\xi$ is a Killing vector, hence $D_\mu\xi_\nu+D_\nu\xi_\mu=0$, reduces relation \eqref{5.6} to
\begin{align}
0&=\xi_{\mu} D_\nu \xi_{\sigma}+\xi_{\nu} D_\sigma \xi_{\mu}+\xi_{\sigma} D_\mu \xi_{\nu}=
\xi_{\mu} D_\nu \xi_{\sigma}-\xi_{\nu} D_\mu \xi_{\sigma}+\xi_{\sigma} D_\mu \xi_{\nu},\notag\\
\xi_{\sigma} (D_\mu \xi_{\nu})&=-(\xi_{\mu} D_\nu \xi_{\sigma}-\xi_{\nu} D_\mu \xi_{\sigma})\label{5.9}.
\end{align}
Contracting equation \eqref{5.9} with $(D^\mu \xi^{\nu})$ yields
\begin{align}
(D^\mu \xi^{\nu})\xi_{\sigma} (D_\mu \xi_{\nu})&=-2(D^\mu \xi^{\nu})(\xi_{\mu} D_\nu \xi_{\sigma}),\notag\\
&=-2\kappa\xi^\nu D_\nu\xi_\sigma,\notag\\
&=-2\kappa^2\xi_\sigma,
\end{align}
where \eqref{5.2} was used. The first property of $\kappa$ is obtained,
\begin{eqnarray}\label{5.13}
\fbox{$
\kappa^2=-\frac{1}{2}(D^\mu \xi^{\nu}) (D_\mu \xi_{\nu})
$}.
\end{eqnarray}
\item
To prove the second property of $\kappa$, the definition of the Riemann tensor is used, i.e. the commutator of two covariant derivative acting on a vector field,
\begin{eqnarray}\label{5.14}
[D_\mu,D_\nu]v_\sigma=R_{\mu\nu\sigma}^{\phantom{\mu\nu\sigma}\kappa}v_\kappa,\hspace{1cm}\forall v_\sigma.
\end{eqnarray}
If the vector field is a Killing vector, i.e. $v_\sigma=\xi_\sigma$, then \eqref{5.14} becomes
\begin{eqnarray}
D_\mu D_\nu\xi_\sigma+D_\nu D_\sigma \xi_\mu = R_{\mu\nu\sigma}^{\phantom{\mu\nu\sigma}\kappa}\xi_\kappa\label{5.15}.
\end{eqnarray}
Writing \eqref{5.15} three times by permuting the indices yields the relation
\begin{eqnarray}\label{5.16}
2D_\mu D_\nu \xi_\sigma =(R_{\mu\nu\sigma}^{\phantom{\mu\nu\sigma}\kappa}-R_{\nu\sigma\mu}^{\phantom{\mu\nu\sigma}\kappa}+R_{\sigma\mu\nu}^{\phantom{\mu\nu\sigma}\kappa})
\xi_\kappa.
\end{eqnarray}
Finally the Bianchi identity $R_{[\mu\nu\sigma]}^{\phantom{[\mu\nu\rho]}\kappa}=0$ allows to write
\begin{eqnarray}
R_{\mu\nu\sigma}^{\phantom{\mu\nu\sigma}\kappa}+R_{\sigma\mu\nu}^{\phantom{\mu\nu\sigma}\kappa}=-R_{\nu\sigma\mu}^{\phantom{\mu\nu\sigma}\kappa},\notag
\end{eqnarray}
and this relation simplifies \eqref{5.16} to $D_\mu D_\nu \xi_\sigma =-R_{\nu\sigma\mu}^{\phantom{\mu\nu\sigma}\kappa}\xi_\kappa$. Equivalently,
\begin{eqnarray}\label{5.18}
\fbox{$
D_\mu D_\nu \xi_\sigma =R_{\sigma\nu\mu}^{\phantom{\mu\nu\sigma}\kappa}\xi_\kappa
$}.
\end{eqnarray}

\item The first two properties are needed to prove that the surface gravity $\kappa$ is constant on the horizon, or more precisely on the orbits of the vector field $\xi$ \cite{Townsend}.

The proof goes as follow. Let us consider  $t$, a tangent vector field to some null hypersurface $\mathcal N$. From the first property \eqref{5.13}, we have
\begin{eqnarray}\label{5.19}
t^\sigma D_\sigma \kappa^2=-(D^\mu \xi^{\nu}) t^\sigma D_\sigma(D_\mu \xi_{\nu}).
\end{eqnarray}
By hypothesis $\xi$ is a null Killing vector, so it is both normal and tangent to the hypersurface $\mathcal N$, so $t^\sigma$ can be chosen as $t^\sigma=\xi^\sigma$ and equation \eqref{5.19} reduces to
\begin{eqnarray}
\xi^\sigma D_\sigma \kappa^2=-(D^\mu \xi^{\nu}) \xi^\sigma D_\sigma D_\mu \xi_{\nu}
=-(D^\mu \xi^{\nu}) \xi^\sigma R_{\mu\nu\sigma\kappa} \xi^{\kappa}=0.
\end{eqnarray}
Where the second property \eqref{5.18} was used and also the fact that the sum over indices $\sigma,\kappa$ that are symmetric in the $\xi$'s while antisymmetric in the Riemann is zero. The surface gravity is thus constant on the orbits of $\xi$. This is the zeroth law of black holes dynamics.

\item To explain the name of $\kappa$, let us consider a particle on a timelike orbit of a Killing field $\xi$. The worldline of the particle is $x^\mu=x^\mu(\lambda)$. Because the particle is moving along an orbit of a Killing field, the four-velocity $u^\mu$ is proportional to $\xi^\mu$ and the factor of proportionality is fixed by requiring the normalization $u^2=u_\mu u^\mu=-1$,
\begin{eqnarray}
u^\mu=A\xi^\mu\Rightarrow u^2=-1=A^2\xi^\mu\xi^\nu g_{\mu\nu}\Rightarrow A^2=\frac{-1}{\xi^2}.\notag
\end{eqnarray}
The four-velocity is thus given by $u^\mu=\frac{\xi^\mu}{\sqrt{-\xi^2}}$. The four-acceleration of the particle along the orbit is
\begin{eqnarray}
a^\mu=\frac{Du ^\mu}{d\lambda}=u^\sigma D_\sigma u^\mu=\frac{\xi^\sigma}{\sqrt{-\xi^2}}D_\sigma\frac{\xi^\mu}{\sqrt{-\xi^2}},\notag
\end{eqnarray}
but $\xi^\sigma D_\sigma \xi^2=2\xi^\sigma\xi^\nu D_\sigma\xi_\nu=0$, because of the antisymmetry of the indices in the two last factors, and the symmetry of them in the to first factors so that the total sum is zero. The acceleration is thus given by
\begin{eqnarray}\label{5.23}
\fbox{$
a^\mu=\frac{\xi^\sigma D_\sigma \xi^\mu}{-\xi^2}
$}.
\end{eqnarray}
On the horizon $-\xi^2$ is nothing else but the square of the red-shift factor. For instance in the case of the Schwarzschild black hole in ingoing Eddigton-Finkelstein coordinates,
\begin{eqnarray}
ds^2=-\left(1-\frac{2M}{r}\right)dv^2+2dvdr+r^2d\Omega^2,
 \end{eqnarray}
$\partial_v$ is a (timelike) Killing vector and we have $-\xi^2=-g_{vv}=\left(1-\frac{2M}{r}\right)$.\\
The relation \eqref{5.2} allows to rewrite the acceleration \eqref{5.23} as
$
a^\mu=\frac{\kappa \xi^\mu}{-\xi^2}$,
so we have
\begin{eqnarray}
\fbox{$
\kappa=\lim_H Va
$},
\end{eqnarray}
with $V=\sqrt{-\xi^2}, a=\sqrt{-a_\mu a^\mu}$. This result means that $\kappa$ is the limiting acceleration when the particle approaches the horizon, hence the name of surface gravity.

\item Finally, the relation \eqref{rel} between the affine distance and the surface gravity is established as follow. Consider the Killing vector $\xi$ of a Killing horizon, $\xi=\frac{\partial}{\partial\alpha}$ for some curve $\alpha=\alpha(\lambda)$ with affine parameter $\lambda$. Thus,
\begin{eqnarray*}
\xi&=\frac{\partial\lambda}{\partial\alpha}\frac{\partial}{\partial\lambda}\equiv fl\hspace{1cm}\text{with}\hspace{1cm}
f=\frac{\partial\lambda}{\partial\alpha},l=\frac{\partial}{\partial\lambda}.
\end{eqnarray*}
The surface gravity is defined by $\kappa=\frac{\partial\ln f}{\partial\alpha}$ and is constant on the horizon. From the definition we get $ \ln f=\kappa\alpha+\ln\kappa$ where the integration constant is chosen to be $\ln\kappa$, so that $f=\kappa e^{\kappa\alpha}$. But $f$ is defined by $f=\frac{\partial\lambda}{\partial\alpha}$, hence the relation between the affine distance $\lambda$ and the surface gravity $\kappa$ is
\begin{eqnarray}
\lambda=e^{\kappa\alpha}.
\end{eqnarray}
In \eqref{rel} the affine parameter was defined to be $\lambda=\epsilon$ and moreover $\alpha=-u$.
\end{enumerate}

\subsubsection{The laws}
%-----------------------------
The result of Hawking shows that the surface gravity plays exactly the same role as the temperature. For completeness, let us just compare the other laws of thermodynamics and the laws of stationary black holes, see \cite{Dowker} for further details and comments:\\
\begin{table}[h]
\centering
\begin{tabular}[c]{c|c|c}
Law &	Thermodynamics	&	Black Hole dynamics\\
\hline
 $0$	& 	$T=cst$					&	$\kappa=cst$\\
 $1$	&	$dE=TdS-pdV$			&	$dM=\frac{\kappa}{8\pi}dA+\Omega dJ$\\
 $2$	&	$dS\ge 0$				&	$dA\ge0$\\
 $3$	&	$T=0$ cannot be reached	&	$\kappa=0$ cannot be reached.
\end{tabular}
\caption{The laws.}
\label{aze}
\end{table}\\
Once again, Hawking's result is the proof that there is not only an analogy between thermodynamics and black holes, but that black holes are indeed thermal objects. Of course nowadays from the AdS/CFT perspective \cite{Aharony}, it is known that a black hole in the gravity side corresponds to a thermal CFT in the field theory side and so, from this duality point of view, Hawking's result concerning the thermal nature of black holes is not so surprising.

\subsection{Information loss paradox and black hole complementarity}
%-------------------------------------------------------------------------------------
This short section aims to provide a very quick overview of the information loss paradox and black hole complementarity. Precise details are beyond the scope of this lecture, but interested readers are invited to consult references \cite{HawkingB} and \cite{Susskind}.
\begin{figure}[!th]
       \centering
      \includegraphics[width=6cm]{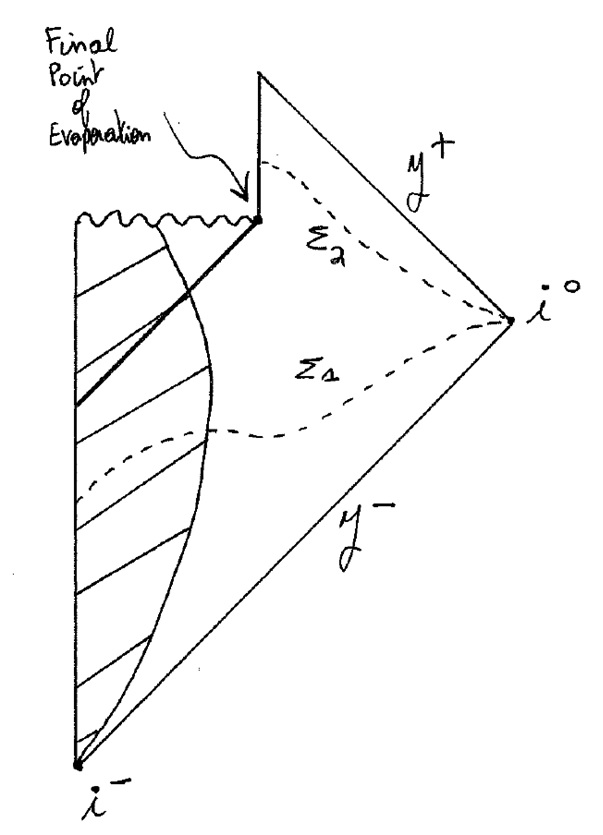}
     \caption{Information loss paradox.}
    \label{fig12}
\end{figure}

After a black hole forms by gravitational collapse, it Hawking radiates and possibly evaporates, after which the spacetime has no longer event horizons.
The situation can be described  by the Carter-Penrose diagram of figure \ref{fig12}.

After the evaporation of the black hole, it is no longer possible to find a Cauchy surface in the spacetime (in figure \ref{fig12} $\Sigma_2$ is not Cauchy for instance), which means that the evolution is non-unitary. Intuitively, whatever is thrown away in the black hole before the evaporation, the situation after which will lead to the same result, independently of what was thrown inside.
Classically, information is thus lost, destroyed by the singularity inside the black hole.

The black hole complementarity is a conjectured solution to this paradox of information loss, according to which the information is both reflected at the horizon and passes through it without being able to escape. The idea is that no observer is able to confirm both stories simultaneously.  The authors of \cite{Susskind} postulated a stretched horizon, a kind of membrane hovering outside the event horizon. The point is that, according to an infalling observer, nothing special happens at the horizon and in this case the information and the infalling observer hit the singularity. But, according to an exterior observer, infalling information heats up the stretched horizon, which then re-radiates it as Hawking radiation.
This is not to say that one has two copies of information (one copies goes inside, the other being Hawking radiated) because this situation is not allowed due to the non-cloning theorem. What can be done is to detect information either inside the black hole or outside, but not both. In this sense, complementarity is to be taken in the quantum sense (non-commuting observer). See \cite{HawkingB} \cite{Susskind} for further details.

\section*{Acknowledgements}
\label{sec:acknowledgements}

\addcontentsline{toc}{section}{Acknowledgments}

It is a pleasure to thank the participants and the other speakers of the ninth Modave summer school in mathematical physics for various interesting discussions and for the relaxing atmosphere they contributed to create during this summer school. In particular, the author warmly thanks Laura Donnay for a critical reading of these notes and for having pointed out misprints in the first version of it. P.-H.L.~benefits from a PhD fellowship of the ULB. This work
is supported in part by the Fund for Scientific Research-FNRS (Belgium), by IISN-Belgium and by ``Communaut\'e fran\c caise de Belgique - Actions de Recherche Concert\'ees''.


\begin{thebibliography}{99}
\bibitem{HawkingP}
S. Hawking, \emph{Particle Creation by Black Hole}, Commun. Math. Phys. 43, 199-220 (1975).


\bibitem{Birrell}
N. Birrell, P. Davies, \emph{Quantum fields in curved space},  Cambridge University Press, Cambridge (1984)

\bibitem{Wald}
R. Wald, \emph{Quantum Field Theory in Curved Spacetime and Black Hole Thermodynamics}, The University of Chicago Press, Chicago (1994).

\bibitem{Carroll}
S. Carroll, \emph{Spacetime and Geometry: An Introduction to General Relativity}, Addison Wesley (2004).

\bibitem{Townsend}
P. Townsend, \emph{Black Holes}, Preprint \href{http://arxiv.org/abs/gr-qc/9707012}{{\ttfamily gr-qc/9707012}}.


\bibitem{Dowker}
F. Dowker, \emph{Black Holes}, Lectures notes, Imperial College London, available at \url{https://dl.dropboxusercontent.com/u/9717190/bh.pdf}.


\bibitem{Henneaux}
M. Henneaux, \emph{Relativit\'e G\'en\'erale}, Lectures notes, Universit\'e Libre de Bruxelles, unpublished.

\bibitem{Ellis}
S. Hawking and G. Ellis, \emph{The large scale structure of space-time}, Cambridge University Press, Cambridge (1973).

\bibitem{Isham}
S. Avis, C. Isham and D. Storey, \emph{Quantum field theory in anti-de Sitter space-time}, Phys. Rev D 18, 3565-3576 (1978).

\bibitem{Geroch}
R. Geroch, \emph{Domain of dependence}, J. Math. Phys 11, 437-449 (1970).


\bibitem{Kiefer}
C. Kiefer, \emph{Thermodynamics of black holes and Hawking radiation}, in \emph{Classical and Quantum Black Holes}, IOP (2002).

\bibitem{Wipf}
A. Wipf, \emph{Quantum fields near Black Holes}, in \emph{Black Holes: Theory and Observation}, lectures notes in Physics 514, 385-415 (1998), Preprint  \href{http://arxiv.org/abs/hep-th/9801025}{{\ttfamily hep-th 9801025}}.

\bibitem{Mukhanov}
V. Mukhanov and S. Winitzki, \emph{Introduction to Quantum Effects in Gravity}, Cambridge University Press, Cambridge (2007).

\bibitem{Traschen}
J. Traschen, \emph{An Introduction to Black Hole Evaporation}, 
Preprint \href{http://arxiv.org/abs/gr-qc/0010055}{{\ttfamily qc/0010055}}.

\bibitem{Barnich}
G. Barnich, \emph{Th\'eorie quantique des champs}, Lectures notes, Universit\'e Libre de Bruxelles, available at
\url{http://homepages.ulb.ac.be/~gbarnich/TQC.pdf}.


\bibitem{Landau}
L. Landau and E. Lifshitz, \emph{The Classical Theory of Field}, Butterworth Heinemann.

\bibitem{Aharony}
O.~Aharony, S.~S. Gubser, J.~Maldacena, H.~Ooguri, and Y.~Oz, \emph{Large {N} field
  theories, string theory and gravity}, {\em Phys. Rept.} {\bfseries 323}
  (2000) 183--386,
Preprint \href{http://arXiv.org/abs/hep-th/9905111}{{\ttfamily hep-th/9905111}}.

\bibitem{HawkingB}
S. Hawking, \emph{Breakdown of predicability in gravitational collapse}, Phys. Rev. D 14, 2460-2473 (1976).

\bibitem{Susskind}
L. Susskind, L. Thorlacius and J. Uglum, \emph{The stretched horizon and black hole complementarity}, Phys. Rev D 48, 3743-3761 (1993).

\end{thebibliography}
\end{document}